\documentclass[9pt,twocolumn,twoside]{osajnl}

\journal{pr} 

\setboolean{shortarticle}{false}

\usepackage{lineno}

\newcommand{\cj}{\mathrm{j}}
\newcommand{\e}{\mathrm{e}}
\newcommand{\rev}[1]{\textcolor{black}{#1}}
\newcommand{\secrev}[1]{\textcolor{black}{#1}}

\newcommand{\improve}{{3}}

\title{Automatic Synthesis of Light Processing Functions for Programmable Photonics: Theory and Realization}

\author[1,*]{Zhengqi Gao}
\author[2,3]{Xiangfeng Chen}
\author[1]{Zhengxing Zhang}
\author[1]{Uttara Chakraborty}
\author[2,3]{Wim Bogaerts}
\author[1]{Duane S. Boning}

\affil[1]{Microsystems Technology Laboratories, Electrical Engineering and Computer Science, Massachusetts Institute of Technology, Cambridge MA 02139, USA}
\affil[2]{Ghent University - IMEC, Department of Information Technology, Gent, Belgium}
\affil[3]{Center of Nano- and Biophotonics, Ghent University, Gent, Belgium}

\affil[*]{Corresponding author: Zhengqi Gao}

\begin{abstract}
 Linear light processing functions (e.g., routing, splitting, filtering) are key functions requiring configuration to implement on a programmable photonic integrated circuit (PPIC). In recirculating waveguide meshes (which include loop-backs), this is usually done manually. Some previous results describe explorations to perform this task automatically, but their efficiency or applicability is still limited. In this paper, we propose an efficient method that can automatically realize configurations for many light processing functions on a square-mesh PPIC. At its heart is an automatic differentiation subroutine built upon analytical expressions of scattering matrices, \rev{that} enables gradient descent optimization for functional circuit synthesis. \rev{Similar to the state-of-the-art synthesis techniques, our method can realize configurations for a wide range of light processing functions, and multiple functions on the same PPIC simultaneously.  However, we do not need to separate the functions spatially into different subdomains of the mesh, and the resulting optimum can have multiple functions using the same part of the mesh. Furthermore, compared to non-gradient or numerical differentiation based methods, our proposed approach achieves \improve x time  reduction in computational cost.}
 \end{abstract}

\setboolean{displaycopyright}{true}

\begin{document}


\maketitle

\section{Introduction}\label{sec:intro}

Photonic integrated circuits (PICs) have drawn increasing attention over the past two decades. Their primary goal is to integrate complex manipulation of light (such as routing, filtering, coupling, interfering) onto a single chip~\cite{spd,wim_silicon_challenge,wim_silicon_challenge2}. Today, a PIC is usually designed for one specific application, so that it can be compact and power-efficient~\cite{bogaerts2020programmable}. The design methodology for these chips is 
similar to that of application-specific integrated circuits (ASIC) in the electronic domain, and thus this kind of PIC is usually referred to as an application-specific PIC (ASPIC).

In contrast, another mainstream type of electronic circuit is the field-programmable gate array (FPGA). These circuits are generic in concept, and their functionality is programmed by configuring the on-chip connectivity of the logical building blocks. The photonic counterpart of FPGA, the programmable photonic integrated circuit (PPIC)~\cite{gao2022automatic,Miller2013forward,miller2013selfforward,taballione20188forward,Bandyopadhyay21forward,clements2016optimalforward,Jinguiji2008synthesisforward,bogaerts2020programmable,perez2017multipurpose,chen2020graph,lopez2020auto,zhuang2015programmable}, has been introduced recently based on the idea of run-time manipulation of light after a chip has been fabricated. Such reconfigurability is usually made available by controlling the active components (e.g., optical phase shifts~\cite{bogaerts2020programmable}) with electrical/thermal signals. Due to its programmability, a PPIC is suitable to various applications such as fast prototyping of ASPICs~\cite{bogaerts2020programmable}, building optical neural networks~\cite{shen2017deep}, and processing quantum information~\cite{madsen2022quantum,arrazola2021quantum}.

A PPIC is composed of a mesh of tunable basic units (TBUs)~\cite{perez2017multipurpose}, also called analog optical gates~\cite{bogaerts2020programmable}. The most common implementation of a TBU is a $2\times 2$ Mach–Zehnder interferometer (MZI) circuit~\cite{bogaerts2020programmable,perez2017multipurpose}. Considering the interconnections of TBUs, PPICs can generally be classified into two categories: (i)~forward-only topologies~\cite{Miller2013forward,miller2013selfforward,taballione20188forward,Bandyopadhyay21forward,clements2016optimalforward,Jinguiji2008synthesisforward,sunil2019matrix,saumil2022single,Sunil2022experimentally}, and (ii)~loop-back (recirculating) topologies~\cite{bogaerts2020programmable,perez2017multipurpose,chen2020graph,lopez2020auto,zhuang2015programmable}. In a forward-only PPIC, light propagates in one direction (e.g., from left to right). It has been proven that with particular forward-only structures, a PPIC can realize any unitary transformation~\cite{PhysRevLett1994unitary_realization,sunil2019matrix,Clements2016optimal}. When fixed-length delay lines are introduced, it is also possible to implement finite impulse response (FIR) digital filters~\cite{Jinguiji2008synthesisforward}. 
\rev{Feed-forward PPICs are commonly used to implement optical neural networks (ONNs)  for AI computing. The first notable experimental realization of an ONN was published in 2017~\cite{shen2017deep}. Later works have considered in-situ ONN training~\cite{saumil2022single,Sunil2022experimentally} via novel optical back-propagation techniques.}

However, without loop-back connections, a forward-only PPIC cannot realize a ring resonator or an infinite impulse response (IIR) digital filter. Such shortcomings have motivated researchers to consider recirculating-based PPICs~\cite{Zhuang15programmable,capmany2016programmable,perez2019scalable}. The most common recirculating configurations are triangular, square, or hexagonal close packing~\cite{Perez16Reconfigurable}. However, while these loop-back meshes offer the possibility to implement more complex connectivities as well as FIR and IIR filters, the configuration of those functions is mostly done by manually assigning and configuring the optical gates in the mesh. Such an ad-hoc method will not be applicable (i)~when we want to synthesize several filters at the same time, and (ii)~when the size of a recirculating PPIC increases substantially.

To address these issues, a few published results~\cite{perez2020multipurpose,lopez2020auto,chen2020graph,perez2020multipurpose2} have proposed methods to perform this task automatically. The authors in~\cite{perez2020multipurpose,perez2020multipurpose2} propose to use optimization techniques to synthesize optical ring resonators and MZIs on a hexagonal-mesh PPIC. In~\cite{lopez2020auto}, the authors propose an auto-routing method based on graph theory for a hexagonal-mesh PPIC, and multi-objective routing is demonstrated by~\cite{chen2020graph}. However, these methods can be dramatically improved to overcome the following key limitations: (i)~their application range is restricted and many light processing functions are not considered; and (ii)~since many optical phase shifts need to be optimized in a PPIC, this high-dimensional optimization problem is not efficient with current methods that rely on non-gradient methods (e.g., PSO in~\cite{perez2020multipurpose}) or gradient methods with numerical differentiation (e.g., Eq. (4)-(5) in the supplementary of~\cite{perez2020multipurpose,perez2020multipurpose2}).


In this paper, we address these two pain points by relying on scattering matrix theory, together with efficient calculation of analytical gradients. Specifically, we propose an efficient method that can realize configurations for many different light processing functions on a square-mesh PPIC, without requiring a priori human design guidance. We start with the compact model of a TBU and derive the analytical transfer functions of the entire circuit according to scattering matrix theory. Built upon this, we implement an automatic differentiation subroutine that can analytically calculate the mean squared error (or other cost functions) between the target frequency responses and the configured circuit responses, and the cost function derivative, with respect to all tunable parameters inside the PPIC. This enables us to efficiently perform gradient descent optimization realizing a variety of light processing functions with different magnitude or/and phase responses. \secrev{Our work has a close relationship with~\cite{perez2019scalable}, where the authors derive a system-level analytical scattering matrix for a hexagonal mesh. However, our approach goes beyond that work by calculating and utilizing gradients for functional synthesis.} In overview, our major contributions include:

\begin{itemize}
\setlength{\itemsep}{1pt}
\setlength{\parsep}{1pt}
\setlength{\parskip}{1pt}
    \item In Section~\ref{sec:theory}, we propose a TBU compact model appropriate for the task of optical filter synthesis, and analytically derive the TBU transfer functions using scattering matrix theory.
    \item In Section~\ref{sec:intuition}, we consider a simplified case where all the horizontal TBUs in a PPIC are fixed to bar states, from which several useful observations can be made.
    \item In Section~\ref{sec:realization}, we demonstrate our efficient synthesis method based on automatic differentiation and gradient descent optimization. We also develop a logarithmic cost function suitable to the case when we want to optimize both the stop band and pass band of a wavelength filter response.
    \item In Section~\ref{sec:result}, we demonstrate that our proposed method can be applied to a wide range of light processing functions at run-time scales of minutes. We also show that our method can synthesize multiple light processing functions simultaneously in the same waveguide mesh.
    \item Finally, in Section~\ref{sec:discussion}, we discuss the limitations of our method and future considerations, such as how to extend it to suit a PPIC containing hundreds or even thousands of TBUs with arbitrary connections.
\end{itemize}

\section{Theory of Scattering Matrices}\label{sec:theory}

\begin{figure}[!hbt]
    \centering
    \includegraphics[width=0.9\linewidth]{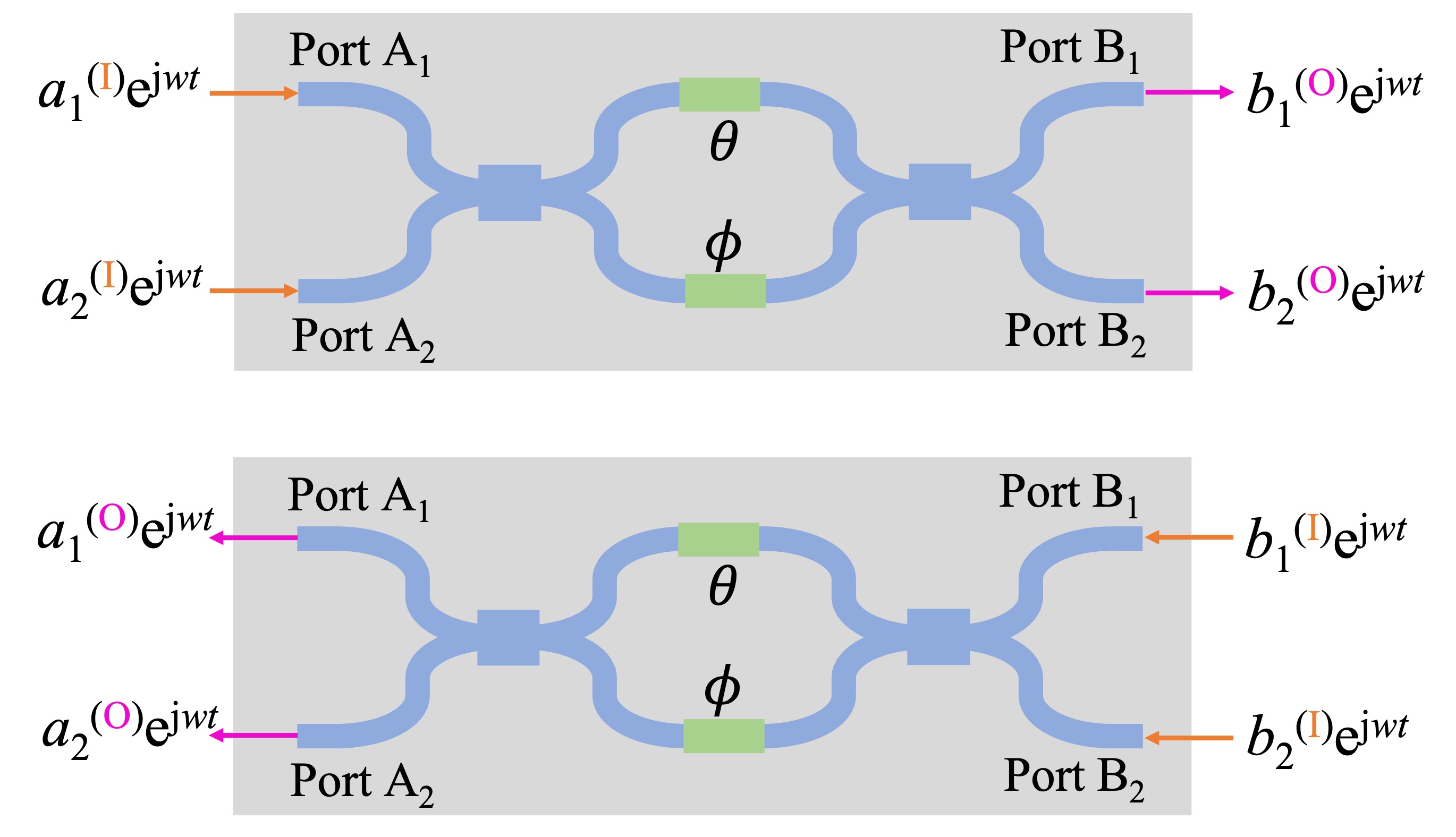}
    \caption{A simplified schematic of a tunable basic unit (TBU). It is made up of two 50\%:50\% directional couplers (DC) on the left and right, and two optical phase shifts (PS) parameterized by $\{\theta,\phi\}$ in the middle. $\{\theta,\phi\}$ can be adjusted freely in $[0,2\pi)$ by thermo- or electro-optic control of the two phase shifts.}
    \label{fig:tbu}
\end{figure}

Following~\cite{bogaerts2020programmable,zhuang2015programmable,Bandyopadhyay21forward}, we consider the TBU structure as shown in Fig.~\ref{fig:tbu} throughout this paper. As shown in the top row of Fig.~\ref{fig:tbu}, we assume that two time-harmonic optical inputs $\{a_1^{(I)}\e^{\cj\omega t},a_2^{(I)}\e^{\cj\omega t}\}$ are provided respectively at the two left ports $\{\text{A}_1,\text{A}_2\}$. Then the outputs can be calculated based on the transfer matrix $\mathbf{F}$:
\begin{equation}\label{eq:basic_Smatrix_input_output}
    \left[
    \begin{array}{c}
         b_1^{(O)} \\
         b_2^{(O)}
    \end{array}\right]=\mathbf{F}
    \left[
    \begin{array}{c}
         a_1^{(I)} \\
         a_2^{(I)}
    \end{array}\right]
\end{equation}
where we use the superscripts `I' and `O' in parentheses to represent the direction of light going into and coming out of the TBU, respectively. The transfer matrix $\mathbf{F}$ is given as~\cite{bogaerts2020programmable,Bandyopadhyay21forward}:
\begin{equation}\label{eq:form_of_Smatrix}
    \mathbf{F} = \underbrace{\frac{\sqrt{2}}{2}\left[
    \begin{array}{cc}
        1 &  -\cj \\
        -\cj &  1
    \end{array}
    \right]}_{\text{right DC}}
    \underbrace{\left[
    \begin{array}{cc}
        \e^{-\cj\theta} &  0 \\
        0 &  \e^{-\cj\phi}
    \end{array}
    \right]}_{\text{PSs}}\underbrace{\frac{\sqrt{2}}{2}\left[
    \begin{array}{cc}
        1 &  -\cj \\
        -\cj &  1
    \end{array}
    \right]}_{\text{left DC}}
\end{equation}
where the optical phases shifts (PSs) are parameterized by $\theta$ and $\phi$, and the directional couplers (DCs) are fixed 50\%:50\% splitters. 
Here we emphasize that the signals $\{a_1^{(I)},a_2^{(I)},b_1^{(O)},b_2^{(O)}\}$ are all complex scalar variables. In our notation, we choose $\e^{\cj\omega t}$ dependence instead of $\e^{-\cj\omega t}$, so that it is consistent with the conventional Fourier transform from the time domain to the frequency domain. This results in the minus signs ahead of the complex unit $\cj$ on the right-hand side of Eq.~(\ref{eq:form_of_Smatrix}); however, from the calculation perspective, the alternative representation can be equivalently employed.

If we reverse the direction of light propagation as shown in the bottom row of Fig.~\ref{fig:tbu}, then the vector on the left-hand side of Eq.~(\ref{eq:basic_Smatrix_input_output}) will be $[a_1^{(O)},a_2^{(O)}]^T$, while $[b_1^{(I)},b_2^{(I)}]^T$ will be on the right. Combining the two propagation cases together, we have the scattering matrix relation:
\begin{equation}\label{eq:tbu_scatter}
\left[
\begin{array}{c}
         b_1^{(O)} \\
         b_2^{(O)}\\
         a_1^{(O)} \\
         a_2^{(O)}
\end{array}
\right]
=
\left[
    \begin{array}{cc}
       \mathbf{F}  & \mathbf{0} \\
        \mathbf{0} & \mathbf{F}
    \end{array}
    \right]
    \left[
\begin{array}{c}
         a_1^{(I)} \\
         a_2^{(I)} \\
         b_1^{(I)} \\
         b_2^{(I)} \\ 
\end{array}
\right]
\end{equation}

For our filter synthesis application, the model in Eq.~(\ref{eq:form_of_Smatrix}) is insufficient: we will never obtain a frequency-dependent response using this model, because $\mathbf{F}$ does not rely on the light frequency $\omega$. To remedy this, we modify the previous transfer matrix by taking the role of the TBU waveguides into consideration:
\begin{equation}\label{eq:our_compact_S}
        \mathbf{F} =     \underbrace{0.5\left[
    \begin{array}{cc}
        \e^{-\cj\theta}-\e^{-\cj\phi} &  -\cj\e^{-\cj\theta}-\cj\e^{-\cj\phi} \\
        -\cj\e^{-\cj\theta}-\cj\e^{-\cj\phi} &   -\e^{-\cj\theta}+\e^{-\cj\phi}
    \end{array}\right]}_{\text{Eq.~(\ref{eq:form_of_Smatrix})}} \alpha\e^{-\cj\omega\frac{n_{\text{eff}}L}{c}}
\end{equation}
where $n_{\text{eff}}(\omega)$ is the effective index of the propagating mode, $L$ represents the length of the waveguide in the TBU, $c$ is the speed of light in free space, and $\alpha$ represents the transmission loss introduced by the waveguides and couplers in the TBU. At the device level, there might be several waveguides in one TBU (e.g., between the left DC and the PSs, between the PSs and the right DC). Our compact model in Eq.~(\ref{eq:our_compact_S}) is valid as long as the waveguides are balanced in the upper and lower arm. See Appendix \ref{sec:appendixa} for more details. Moreover, without considering dispersion (i.e., $n_{\text{eff}}$ is a constant independent of $\omega$), the $\e^{-\cj\omega\frac{n_{\text{eff}}L}{c}}$ factor naturally corresponds to a time delay $\frac{n_{\text{eff}}L}{c}$ according to the Fourier transform, and thus PPICs can rely on digital filter theory to realize optical filter functions.

\begin{figure}[!htb]
    \centering
    \includegraphics[width=1.0\linewidth]{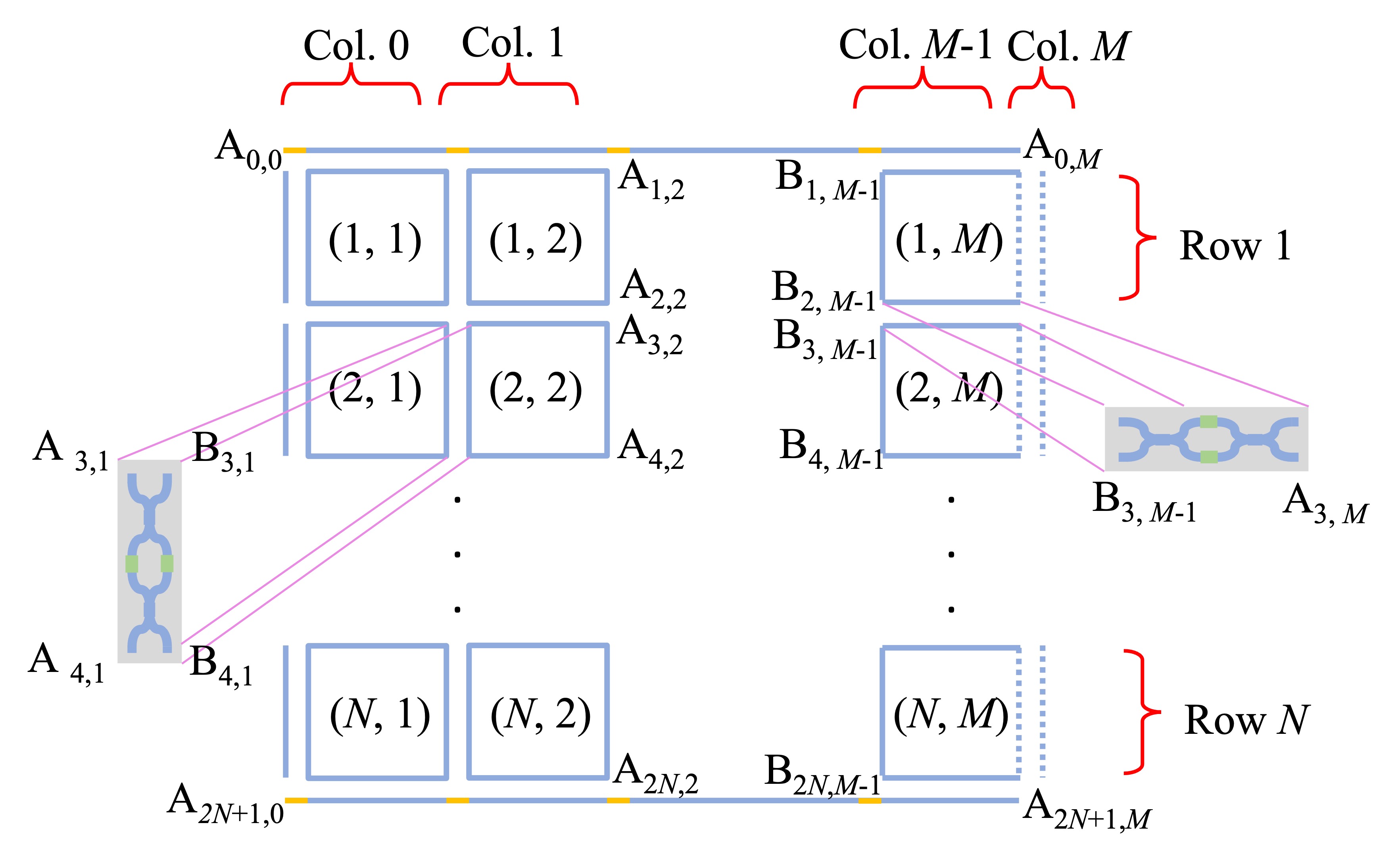}
    \caption{The schematic of an $N\times M$ square-mesh PPIC. For derivation simplicity, we disable the TBUs at the right-most column marked by dashed lines and assume that the top and bottom connections (yellow lines) are ideal.}
    \label{fig:ppic}
\end{figure}

\begin{figure}[!htb]
    \centering
    \includegraphics[width=1.0\linewidth]{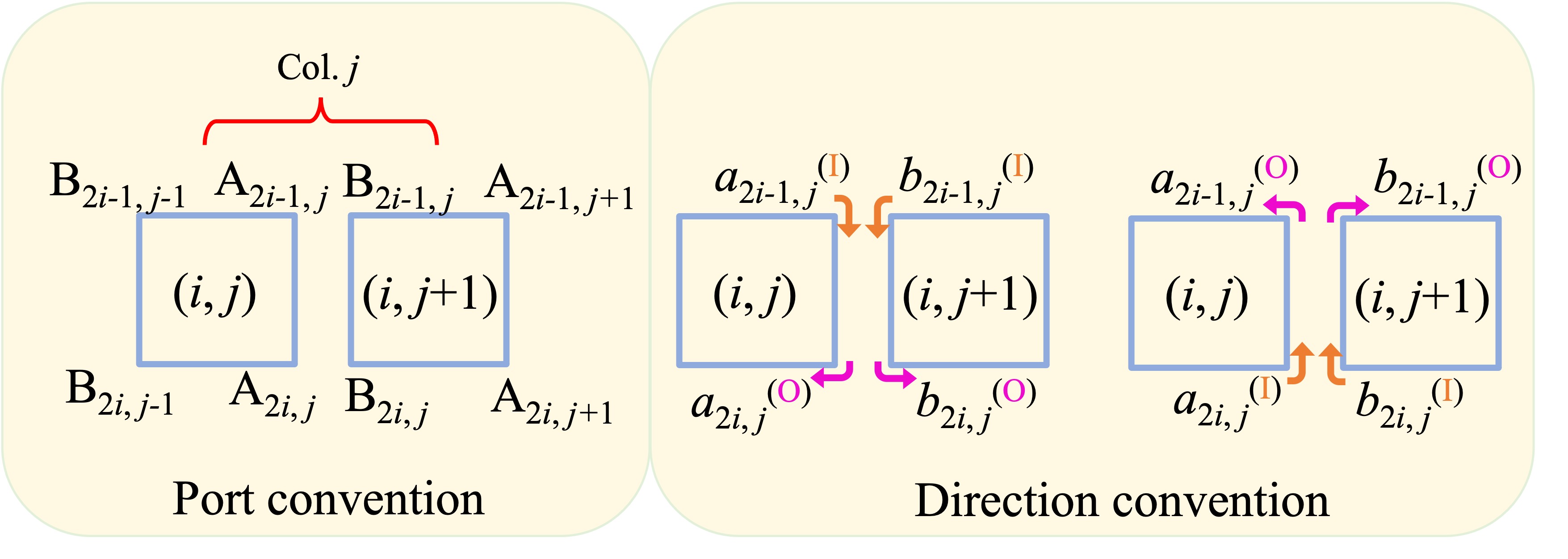}
    \caption{Naming conventions for port and direction. Capitalized `A' and `B' should be regarded as port names, and small `a' and `b' are the complex magnitudes ahead of $e^{j\omega t}$. For conciseness, we have omitted the $e^{j\omega t}$ dependence.}
    \label{fig:ppic_name_convention}
\end{figure}

The circuit schematic of the recirculating PPIC waveguide mesh is shown in Fig.~\ref{fig:ppic}. In this paper, we ignore the TBUs in the right-most column. We also assume that the top and bottom connections (the yellow lines) are ideal connections, i.e., their transfer function is identity. These two assumptions are made for mathematical simplicity and the purpose of demonstration, and note that our method is applicable without these assumptions.

Next, we introduce naming conventions for the ports and propagation directions. As shown in Fig.~\ref{fig:ppic_name_convention}, we adopt the following conventions for the ports in this PPIC: (i)~the letters `A' and `B' are used to denote the ports on the left and right edge of a vertical TBU, respectively. (ii)~The subscript $(n,m)$ is used to express that the port is on the $n$-th row and $m$-th column, where $n=0,1,\cdots,2N+1$ and $m=0,1,\cdots,M$. As shown in Fig.~\ref{fig:ppic_name_convention}, for any port, the light can propagate in two directions. We define going into and out of the vertical TBU device as `I' (i.e., orange arrows) and `O' (i.e., purple arrows), respectively. One minor subtlety arises when applying this direction naming convention to the top line shown in Fig.~\ref{fig:ppic}, since this top line does not associate with any vertical device. In this case, we consider there to be virtual vertical TBUs above this top line, and then apply our direction naming convention. Similarly, we consider there to be virtual vertical TBUs beneath the bottom line in Fig.~\ref{fig:ppic} for the purpose of notation consistency.

Following Eq.~(\ref{eq:tbu_scatter}) and applying the scattering matrix to the two propagating directions shown in the right figure in Fig.~\ref{fig:ppic_name_convention}, we have:
\begin{equation}\label{eq:vertical_tbu_transfer}
\left[
\begin{array}{c}
     a_{2i,j}^{(O)}\\
     b_{2i,j}^{(O)}\\
     a_{2i-1,j}^{(O)}\\
     b_{2i-1,j}^{(O)}\\
\end{array}
\right]
=
\left[
    \begin{array}{cc}
       \mathbf{F}  & \mathbf{0} \\
        \mathbf{0} & \mathbf{F}
    \end{array}
    \right]
    \left[
\begin{array}{c}
     a_{2i-1,j}^{(I)}\\
     b_{2i-1,j}^{(I)}\\
     a_{2i,j}^{(I)}\\
     b_{2i,j}^{(I)}\\
\end{array}
\right].
\end{equation}
If we expand all terms in Eq.~(\ref{eq:vertical_tbu_transfer}) and rearrange the terms to locate those related to 'b' and 'a' at the left-hand and right-hand side, respectively, we obtain:
\begin{equation}\label{eq:transformed_vertical_transferfunc}
    \left[
\begin{array}{c}
     b_{2i-1,j}^{(I)}\\
     b_{2i-1,j}^{(O)}\\
     b_{2i,j}^{(I)}\\
     b_{2i,j}^{(O)}\\
\end{array}
\right]
=
\mathbf{V}
    \left[
\begin{array}{c}
     a_{2i-1,j}^{(I)}\\
     a_{2i-1,j}^{(O)}\\
     a_{2i,j}^{(I)}\\
     a_{2i,j}^{(O)}\\
\end{array}
\right]
\end{equation}
where $\mathbf{V}$ is of size $4\times 4$, and  with some algebra, we have:
\begin{equation}\label{eq:relation_f_s}
    \begin{aligned}
            V_{11}&=V_{33}= -F_{11}/F_{12}\\
            V_{22}&=V_{44}=F_{22}/F_{12}\\
            V_{41}&=V_{32}=1/F_{12}\\
            V_{23}&=V_{41}=F_{21}-F_{11}F_{22}/F_{12}\\
    \end{aligned}
\end{equation}
and other entries of $\mathbf{V}$ are all zero. Here we use $F_{kl}$ to denote the entry on the $k$-th ($k=1,2$) row and $l$-th ($l=1,2$) column of matrix $\mathbf{F}$. Similar notations are applied to $\mathbf{V}$ as well as all later occurring matrices. It is important to note that Eq.~(\ref{eq:transformed_vertical_transferfunc}) holds for the index $ i=1,2,\cdots,N$ and $j=0,2,\cdots,M-1$, which covers all vertical TBUs in the middle, except the top and bottom line in Fig.~\ref{fig:ppic}. The top and bottom lines correspond to row index $0$ and $2N+1$ under our naming convention, and the following relations hold on these two lines because we assume the yellow lines in Fig. ~\ref{fig:ppic} are ideal:
\begin{equation}\label{eq:vertical_boundary}
        \left[
\begin{array}{c}
     b_{k,j}^{(I)}\\
     b_{k,j}^{(O)}\\
\end{array}
\right]
=
\left[
    \begin{array}{cc}
       0  & 1 \\
       1 & 0
    \end{array}
    \right]
    \left[
\begin{array}{c}
     a_{k,j}^{(I)}\\
     a_{k,j}^{(O)}\\
\end{array}
\right]\quad k=0 \text{ or } 2N+1 \; .
\end{equation}

Recall that in writing Eq.~(\ref{eq:vertical_tbu_transfer}), we apply the scattering matrix method to the two propagation directions of a vertical TBU. We can do the same thing for a horizontal TBU, which gives the following equation:
\begin{equation}
    \left[
\begin{array}{c}
     a_{2i,j+1}^{(I)}\\
     a_{2i+1,j+1}^{(I)}\\
     b_{2i,j}^{(I)}\\
     b_{2i+1,j}^{(I)}\\
\end{array}
\right]
=
\left[
    \begin{array}{cc}
       \mathbf{F}  & \mathbf{0} \\
        \mathbf{0} & \mathbf{F}
    \end{array}
    \right]
    \left[
\begin{array}{c}
     b_{2i,j}^{(O)}\\
     b_{2i+1,j}^{(O)}\\
     a_{2i,j+1}^{(O)}\\
     a_{2i+1,j+1}^{(O)}\\
\end{array}
\right] \; .
\end{equation}
Similarly, we now move all terms related to 'b' and 'a' to the right-hand and left-hand side, respectively, and obtain:
\begin{equation}\label{eq:transformed_horizontal_transferfunc}
    \left[
\begin{array}{c}
     a_{2i,j+1}^{(I)}\\
    a_{2i,j+1}^{(O)}\\
     a_{2i+1,j+1}^{(I)}\\
     a_{2i+1,j+1}^{(O)}\\
\end{array}
\right]
=
\mathbf{H}
    \left[
\begin{array}{c}
     b_{2i,j}^{(I)}\\
     b_{2i,j}^{(O)}\\
    b_{2i+1,j}^{(I)}\\
     b_{2i+1,j}^{(O)}\\
\end{array}
\right]
\end{equation}
where $\mathbf{H}$ is of size $4\times 4$ and  with some algebra, we have:
\begin{equation}\label{eq:relation_g_s}
    \begin{aligned}
            H_{12}&=F_{11},&H_{14}&=F_{12}\\
            H_{32}&=F_{21},&H_{34}&=F_{22}\\
            H_{21}&=F_{22}/\text{det}(\mathbf{F}),&H_{23}&=-F_{12}/\text{det}(\mathbf{F})\\
            H_{41}&=-F_{21}/\text{det}(\mathbf{F}),&H_{43}&=F_{11}/\text{det}(\mathbf{F})\\
    \end{aligned}
\end{equation}
and all other entries of $\mathbf{H}$ are zero. Here  $\text{det}(\mathbf{F})=F_{11}F_{22}-F_{12}F_{21}$ represents the determinant of $\mathbf{F}$. Again note that Eq.~(\ref{eq:transformed_horizontal_transferfunc}) holds for the index $i=0,2,\cdots,N$ and $j=0,2,\cdots,M-1$.

For a specific column index $j$, if we vary the row index $i$ and $k$ in Eqs.~(\ref{eq:transformed_vertical_transferfunc}) and (\ref{eq:vertical_boundary}), and next stack all the resulting equations in one column, we obtain a scattering matrix for the mapping: $\{a_{n,j}^{(I)},a_{n,j}^{(O)}\}{\rightarrow}\{b_{n,j}^{(I)},b_{n,j}^{(O)}\}$,  where $n=0,1,\cdots,2N+1$. Similarly, if we vary the row index $i$ in Eq.~(\ref{eq:transformed_horizontal_transferfunc}),  we can write down the scattering matrix for the mapping: $\{b_{n,j}^{(I)},b_{n,j}^{(O)}\}{\rightarrow}\{a_{n,j+1}^{(I)},a_{n,j+1}^{(O)}\}$. Combining these two steps together gives us the scattering matrix representing the mapping: $\{a_{n,j}^{(I)},a_{n,j}^{(O)}\}\rightarrow\{a_{n,j+1}^{(I)},a_{n,j+1}^{(O)}\}$. Mathematically, that is to say:
\begin{equation}\label{eq:the_j_column}
    \left[
\begin{array}{c}
     a_{0,j+1}^{(I)}\\
    a_{0,j+1}^{(O)}\\
     \vdots\\
     a_{2N+1,j+1}^{(I)}\\
    a_{2N+1,j+1}^{(O)}\\
\end{array}
\right]
=
\mathbf{T}^j
    \left[
\begin{array}{c}
     a_{0,j}^{(I)}\\
    a_{0,j}^{(O)}\\
     \vdots\\
     a_{2N+1,j}^{(I)}\\
    a_{2N+1,j}^{(O)}\\
\end{array}
\right]
\end{equation}
where $\mathbf{T}^j$ is of size $(4N+4)\times(4N+4)$ and can be expressed as the product of two block diagonal matrices:
\begin{equation}\label{eq:expression_j_layer_transfer}
\begin{aligned}
        \mathbf{T}^j=&\text{Diag}(\overbrace{\mathbf{H},\cdots,\mathbf{H}}^{(N+1) })\\
        &\times\text{Diag}(\left[
    \begin{array}{cc}
       0  & 1 \\
       1 & 0
    \end{array}
    \right],\underbrace{\mathbf{V},\cdots,\mathbf{V}}_{N },\left[
    \begin{array}{cc}
       0  & 1 \\
       1 & 0
    \end{array}
    \right])
\end{aligned}
\end{equation}
We emphasize that the first block diagonal matrix on the right-hand side of Eq.~(\ref{eq:expression_j_layer_transfer}) is constructed via putting $(N+1)$ $\mathbf{H}$ matrices along the main diagonal. Here readers should be aware that the $(N+1)$ $\mathbf{H}$ matrices correspond to $(N+1)$ different horizontal TBUs from top to bottom, and that to keep the notation uncluttered, we have not introduced subscripts or superscripts on $\mathbf{H}$ to distinguish 
them. Each $\mathbf{H}$ might be different. The second matrix on the right-hand side of Eq.~(\ref{eq:expression_j_layer_transfer}) is constructed by putting a $2\times 2$ matrix at the front and end, while the middle is filled with $N$ $\mathbf{V}$ matrices corresponding to $N$ different vertical TBUs from top to bottom. Similarly, each $\mathbf{V}$ can be different. 

If we repeat Eq.~(\ref{eq:the_j_column}) $M$ times for different column index $j$, then we can obtain the overall scattering matrix for the mapping from $\{a_{n,0}^{(I)},a_{n,0}^{(O)}\}$ to $\{a_{n,M}^{(I)},a_{n,M}^{(O)}\}$:

\begin{equation}
        \left[
\begin{array}{c}
     a_{0,M}^{(I)}\\
    a_{0,M}^{(O)}\\
     \vdots\\
     a_{2N+1,M}^{(I)}\\
    a_{2N+1,M}^{(O)}\\
\end{array}
\right]
=
\mathbf{T}
    \left[
\begin{array}{c}
     a_{0,0}^{(I)}\\
    a_{0,0}^{(O)}\\
     \vdots\\
     a_{2N+1,0}^{(I)}\\
    a_{2N+1,0}^{(O)}\\
\end{array}
\right]
\end{equation}
where 
\begin{equation}\label{eq:t}
    \mathbf{T}=\mathbf{T}^{M-1}\cdots\mathbf{T}^1\mathbf{T}^0 \; .
\end{equation}
If we rearrange the order of entries to put the values related to the direction `I' in the first several rows, and those related to the direction `O' in the last several rows, we obtain:
\begin{equation}\label{eq:overall_transfer_func}
        \left[
\begin{array}{c}
     a_{0,M}^{(I)}\\
     \vdots\\
    a_{2N+1,M}^{(I)}\\
         a_{0,M}^{(O)}\\
    \vdots\\
    a_{2N+1,M}^{(O)}\\
\end{array}
\right]
=
\mathbf{P}^T\mathbf{T}\mathbf{P}
    \left[
\begin{array}{c}
     a_{0,0}^{(I)}\\
     \vdots\\
    a_{2N+1,0}^{(I)}\\
    a_{0,0}^{(O)}\\
    \vdots\\
    a_{2N+1,0}^{(O)}\\
\end{array}
\right]
\end{equation}
where $\mathbf{P}$ is a row permutation matrix of size $(4N+4)\times(4N+4)$. Note that $\mathbf{P}$ has a known structure and its entries are either $0$ or $1$. For later simplicity, we will introduce the symbol $\mathbf{a}_{M}^{(I)}=[a_{0,M}^{(I)},\cdots,a_{2N+1,M}^{(I)}]^T$ and $\mathbf{a}_{M}^{(O)}=[a_{0,M}^{(O)},\cdots,a_{2N+1,M}^{(O)}]^T$ for the left-hand side of Eq.~(\ref{eq:overall_transfer_func}). Similar notations are also used for the right-hand side of Eq.~(\ref{eq:overall_transfer_func}), so that it can be simplified as:
\begin{equation}\label{eq:final_transfer_short}
        \left[
\begin{array}{c}
\mathbf{a}_{M}^{(I)}\\
\mathbf{a}_{M}^{(O)}
\end{array}
\right]
=
\mathbf{T}^\star
    \left[
\begin{array}{c}
\mathbf{a}_{0}^{(I)}\\
\mathbf{a}_{0}^{(O)}
\end{array}
\right]=\left[\begin{array}{cc}
    \mathbf{T}^\star_{11} & \mathbf{T}^\star_{12} \\
     \mathbf{T}^\star_{21} & \mathbf{T}^\star_{22}
\end{array}
\right]\left[
\begin{array}{c}
\mathbf{a}_{0}^{(I)}\\
\mathbf{a}_{0}^{(O)}
\end{array}
\right]
\end{equation}
where
\begin{equation}\label{eq:tstar}
    \mathbf{T}^\star=\mathbf{P}^T\mathbf{T}\mathbf{P}
\end{equation}
and we adopt the block matrix notation in the last equality. 

Thus far, we have obtained a relation between the input and the output. The ultimate scattering matrix $\mathbf{T}^\star$ is related to the individual $\mathbf{V}$ (or $\mathbf{H}$) matrix of a vertical (or horizontal) TBU device via Eqs.~(\ref{eq:tstar}), (\ref{eq:t}), and (\ref{eq:expression_j_layer_transfer}) in sequence. Furthermore, the relation from $\mathbf{V}$ and $\mathbf{H}$ matrices to the individual phase shifts $\{\theta,\phi\}$ are also clear via Eqs.~(\ref{eq:relation_g_s}), (\ref{eq:relation_f_s}), and (\ref{eq:our_compact_S}). Thus, we have obtained an analytical expression of $\mathbf{T}^\star$ defined by all phase shifts $\{\theta,\phi\}$. Although it is difficult to explicitly write down the expression for every entry in the $\mathbf{T}^\star$ matrix, we do know the sequential operations to construct it. Most importantly, all of the operations involved (e.g., matrix-vector multiplication) are differentiable, so that we can easily calculate $\frac{\partial \mathbf{T}^\star}{\partial \theta}$ or $\frac{\partial \mathbf{T}^\star}{\partial \phi}$ for any $\{\phi,\theta\}$ of any TBU device. As demonstrated later, this will form the basis for our synthesis method.

\begin{figure}[!htb]
    \centering
    \includegraphics[width=1.0\linewidth]{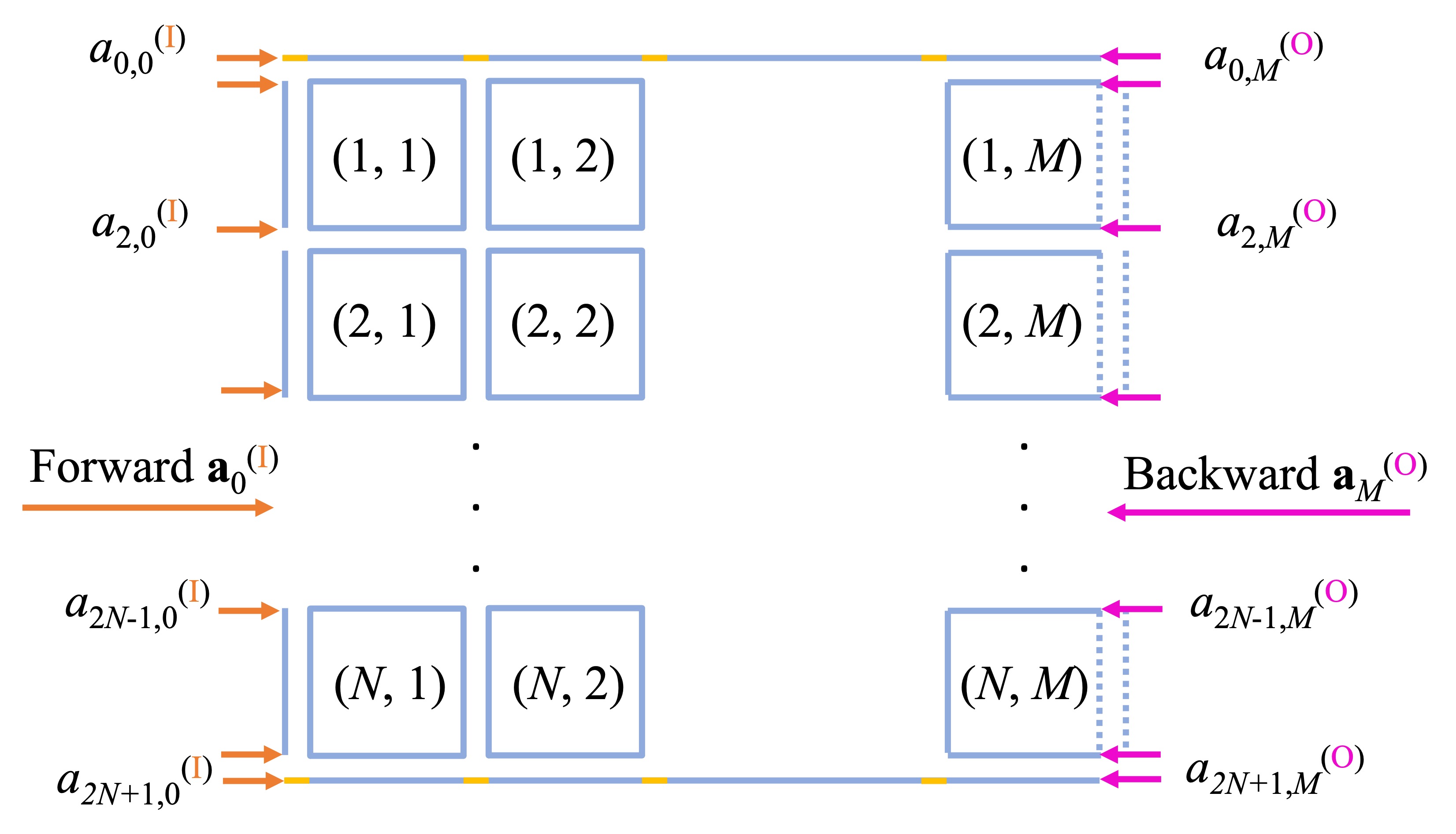}
    \caption{Illustration of the forward input at the left and the backward input at the right. Note that the directions `I' and `O' in the parenthesized superscripts are defined according to going into or coming out of the associated vertical TBU as defined in Fig.~\ref{fig:ppic_name_convention}.}
    \label{fig:ppic_inputoutput}
\end{figure}

Without loss of generality, we assume that our desired forward light propagation is from left to right in the PPIC shown in Fig.~\ref{fig:ppic_inputoutput}. Then we can regard the forward input $\mathbf{a}_{0}^{(I)}$ at the left and the backward input $\mathbf{a}_{M}^{(O)}$ at the right both as  given constant vectors. Based on Eq.~(\ref{eq:final_transfer_short}), we can now express the forward output at the right $\mathbf{a}_{M}^{(I)}$ and the backward output at the left $\mathbf{a}_{0}^{(O)}$ as:
\begin{equation}\label{eq:expression_of_four_ports}
    \begin{aligned}
            \mathbf{a}_{M}^{(I)}&=(\mathbf{T}_{11}^\star-\mathbf{T}_{12}^\star\mathbf{T}_{22}^{\star,-1}\mathbf{T}_{21})\mathbf{a}_{0}^{(I)}+\mathbf{T}_{12}\mathbf{T}_{22}^{\star,-1}\mathbf{a}_{M}^{(O)}\\
            \mathbf{a}_{0}^{(O)}&=\mathbf{T}_{22}^{\star,-1}\mathbf{a}_{M}^{(O)}-\mathbf{T}_{22}^{\star,-1}\mathbf{T}^\star_{21}\mathbf{a}_{0}^{(I)}\\
    \end{aligned}
\end{equation}
where we use $\mathbf{T}_{22}^{\star,-1}$ to represent the inverse of the matrix $\mathbf{T}_{22}^\star$. In reality, the backward input at right $\mathbf{a}_{M}^{(O)}$ will usually be set to a zero vector, and the forward output at right is regarded as the final response of the PPIC:
\begin{equation}\label{eq:only_consider_forward_output}
    \begin{aligned}
            \mathbf{a}_{M}^{(I)}&=(\mathbf{T}_{11}^\star-\mathbf{T}_{12}^\star\mathbf{T}_{22}^{\star,-1}\mathbf{T}_{21})\mathbf{a}_{0}^{(I)}=\mathbf{G}^\star\mathbf{a}_{0}^{(I)}
    \end{aligned}
\end{equation}
where for simplicity, we have denoted $\mathbf{G}^\star=\mathbf{T}_{11}^\star-\mathbf{T}_{12}^\star\mathbf{T}_{22}^{\star,-1}\mathbf{T}_{21}$.

Several points are worth noting. First, both $\mathbf{a}_{0}^{(I)}$ and  $\mathbf{a}_{M}^{(I)}$ are of size $(2N+2)\times 1$. This provides us some flexibility to synthesize multiple light processing functions simultaneously. For instance, we can feed an input wave from the top port of the first vertical TBU, i.e., ${a}_{1,0}^{(I)}$ equal to $1$ and all other entries of $\mathbf{a}_{0}^{(I)}$ equal to $0$. Then the outputs at the second and third entry of $\mathbf{a}_{M}^{(I)}$ can be used to synthesize two different light processing functions. Second, $\mathbf{a}_{0}^{(O)}$ might not be zero in Eq.~(\ref{eq:expression_of_four_ports}) even if $\mathbf{a}_{M}^{(O)}$ is zero, because the information brought by $\mathbf{a}_{0}^{(I)}$ can recirculate back. This is revealed by the term $\mathbf{T}_{22}^{\star,-1}\mathbf{T}^\star_{21}\mathbf{a}_{0}^{(I)}$ at the second line in Eq.~(\ref{eq:expression_of_four_ports}). Third, recall we assume that the yellow lines in Fig.~\ref{fig:ppic} are ideal connections, leading to the zero-one matrix in Eqs.~(\ref{eq:vertical_boundary}) and (\ref{eq:expression_j_layer_transfer}). If the yellow lines are instead not ideal, we just need to revise the $2\times 2$ zero-one matrix, while our derivation (as well as the later synthesis method) still holds.

We make two additional remarks related to the yellow direct connections in the top and bottom rows. First, from the application perspective, the yellow direct connections in the top and bottom rows of the mesh introduce a peculiarity. These connections break the connection symmetry of the mesh, and in particular break the clockwise/counterclockwise degeneracy of the square waveguide mesh.  Normally, when injecting light in a square waveguide mesh, light will either circulate in a clockwise or counterclockwise direction inside a unit cell, but these circulations are not coupled. This means that in the scattering matrix of a square waveguide mesh, at least half of the elements are zero. By adding the connections in the top and bottom row, these clockwise/counterclockwise circulations can be coupled and more generic mesh functions can be defined. 

Secondly, from the calculation perspective, introducing the yellow direct connections makes us only need to provide the forward input $\mathbf{a}_0^{(I)}$ (i.e., $2N+2$ scalars) if we assume the backward input $\mathbf{a}_M^{O}=\mathbf{0}$. However, without these yellow direct connections, we would have to set input values for those floating ports in the top and bottom row, otherwise the conditions are insufficient to determine the circuit response. Analytical gradients can still be calculated in such case, but our derivation will need substantial modification.

\section{Implications from Horizontal Relaxation}\label{sec:intuition}

In the previous section, we derive the scattering matrix for a square-mesh PPIC in a general form. In this section, we consider a simplified case under the assumption of horizontal relaxation: all horizontal TBUs are configured with $\theta=0$ and $\phi=\pi$, and thus operate in the bar state~\cite{bogaerts2020programmable}. This implies that in Fig.~\ref{fig:tbu}, the light propagates from Port~$\text{A}_1$ to $\text{B}_1$, $\text{A}_2$ to $\text{B}_2$, or reversely, but does not go from $\text{A}_1$ to $\text{B}_2$. Namely, when passing a horizontal TBU, the light is confined in the upper or lower arm.

\begin{figure}[!htb]
    \centering
    \includegraphics[width=0.75\linewidth]{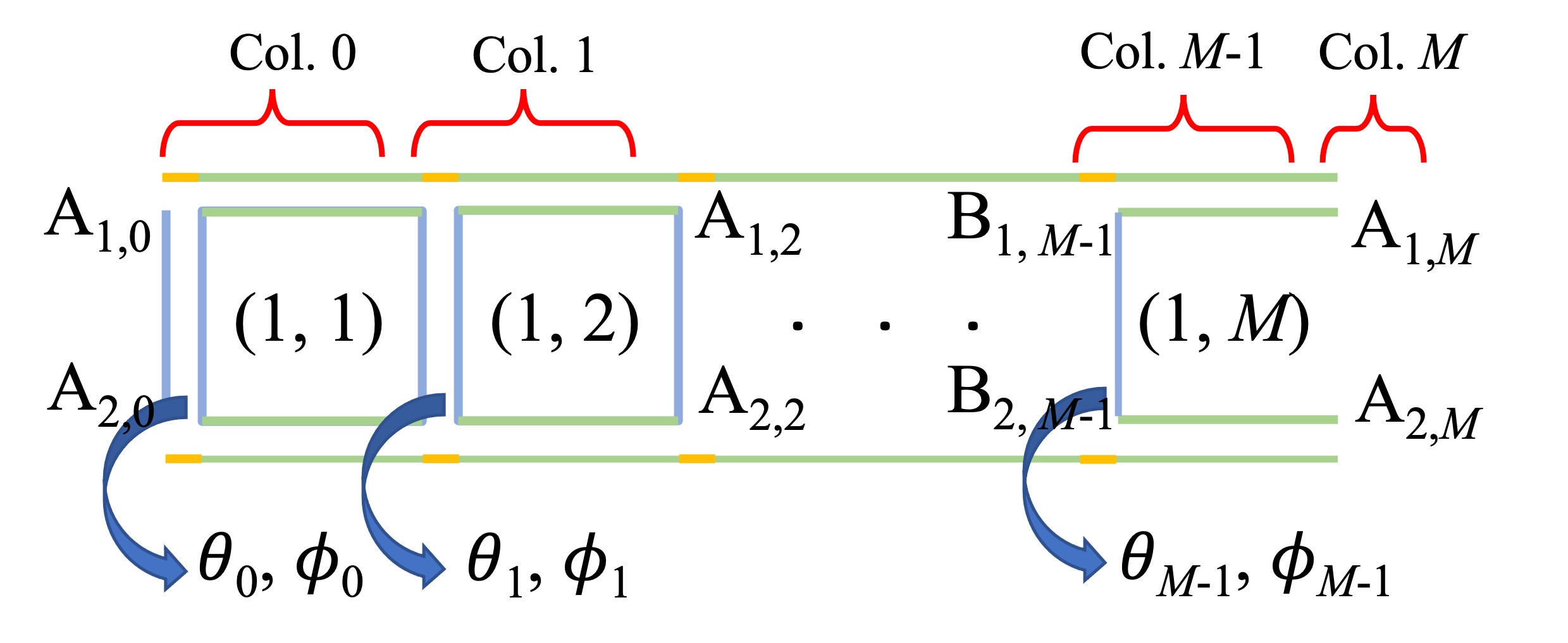}
    \caption{The schematic of a $1\times M$ square PPIC. The phase shifts in green horizontal TBUs are fixed to $\theta=0$ and $\phi=\pi$.}
    \label{fig:ppic1M_horizontalrelax}
\end{figure}

As a starting point, we consider a $1\times M$ square-mesh PPIC under this horizontal relaxation. Its schematic is shown in Fig.~\ref{fig:ppic1M_horizontalrelax}. In this case, the $\mathbf{G}^\star$ matrix defined in Eq.~(\ref{eq:only_consider_forward_output}) is of size $4\times 4$. When $M$ is odd, its expression is:
\begin{equation}\label{eq:hmatrix_in_1M}
     \mathbf{G}^\star=\left[
    \begin{array}{cccc}
      \e^{-\cj M\omega \tau}  &  0             & 0   & 0     \\
        0  & 0        & -\xi_M    & 0     \\
        0  & \xi_M               &  0 & 0  \\
        0  & 0              & 0   & -\e^{-\cj M\omega \tau} \\
    \end{array}
    \right]\,(M=1,3,\cdots)
\end{equation}
and when $M$ is even, its expression is:
\begin{equation}\label{eq:hmatrix_in_1M_even}
              \mathbf{G}^\star=
    \left[
    \begin{array}{cccc}
      \e^{-\cj M\omega \tau}  &  0             & 0   & 0     \\
        0  & \xi_M         & 0   & 0     \\
        0  & 0              &  \xi_M & 0  \\
        0  & 0              & 0   & \e^{-\cj M\omega \tau} \\
    \end{array}
    \right]\, (M=2,4,\cdots) \; .
\end{equation}
The $\e^{-\cj M\omega \tau}$ term inside the matrix corresponds to the output at the top or bottom line in Fig.~\ref{fig:ppic1M_horizontalrelax}. Intuitively, this makes sense, since we have $M$ horizontal TBUs at the top line, and under horizontal relaxation, these $M$ TBUs function as $M$ time delay elements. This in turns implies that the absolute value of $\xi_M$ in Eqs.~(\ref{eq:hmatrix_in_1M}) and (\ref{eq:hmatrix_in_1M_even}) is more interesting, and can be utilized to synthesize light processing functions. $|\xi_M|$ can be proven to have the following form:
\begin{equation}\label{eq:def_xim}
\begin{aligned}
        |\xi_M| = \left|\frac{\prod_{m=0}^{M-1} (c_mf_m-d_me_m)}{\left[
        \begin{array}{cc}
            0 & 1 \\
        \end{array}
        \right]\cdot\prod_{m=0}^{M-1}\left[
    \begin{array}{cc}
        c_m &  d_m\\
        e_m &  f_m 
    \end{array}
    \right]\cdot
    \left[\begin{array}{c}
         0 \\
         1
    \end{array}\right]}\right| \\
\end{aligned}
\end{equation}
where $\{c_m,d_m,e_m,f_m\}$ are scalar values associated with the $m$-th vertical TBU ($m=0,1,\cdots,M-1$). Specifically, as shown in Fig.~\ref{fig:ppic1M_horizontalrelax}, if we denote the phase shifts inside the $m$-th vertical TBU as $\{\theta_m,\phi_m\}$, we have the following relations:
\begin{equation}\label{eq:def_of_cdef}
\begin{aligned}
            c_m&=\cj \e^{-\cj 2\omega \tau}\frac{p_m^2-q_m^2}{2q_m}\\
    d_m&=-\cj\e^{-\cj\omega \tau}\frac{p_m}{q_m}\\
    e_m&=\cj\e^{\cj\omega \tau}\frac{p_m}{q_m}\\
    f_m&=-2\cj\e^{\cj2\omega \tau}\frac{1}{q_m}\\
\end{aligned}
\end{equation}
where for simplicity, we have denoted $\tau(\omega)=\frac{n_{\text{eff}}(\omega) L}{c}$, and:
\begin{equation}\label{eq:def_p_q}
    \begin{aligned}
    p_m &= \e^{-\cj\theta_m} - \e^{-\cj\phi_m} \\
    q_m &= -\e^{-\cj\theta_m} - \e^{-\cj\phi_m} \; .\\
    \end{aligned}
\end{equation}
With Eq.~(\ref{eq:def_of_cdef}), the numerator in Eq.~(\ref{eq:def_xim}) can be proven to be $1$, and the denominator is a polynomial in $\e^{\cj 2\omega t}$. As an example, we have:
\begin{equation}
    \begin{aligned}
    |\xi_1| &=|\frac{1}{2\e^{\cj 2\omega \tau}/q_0}|=|\frac{q_0}{2}\e^{-\cj 2\omega \tau}|\\
    |\xi_2| &= |\frac{1}{(-4\e^{\cj 4\omega \tau}+p_0p_1)/q_0q_1}|=|\frac{q_0q_1\e^{-\cj 4\omega \tau}}{-4+p_0p_1\e^{-\cj 4\omega \tau}}|\\
    \end{aligned}
\end{equation}
Thus, a $1\times M$ square PPIC, under horizontal relaxation, can be used to synthesize an IIR filter with zeros at the origin and poles generally complex.

A natural thought would be to extend the $1\times M$ square PPIC under horizontal relaxation to an $N\times M$ square PPIC. Fortunately, due to the assumptions of our horizontal relaxation, this is straightforward. Specifically, in Eqs.~(\ref{eq:hmatrix_in_1M})-(\ref{eq:hmatrix_in_1M_even}), we have a $\mathbf{G}^\star$ matrix with size of $4\times 4$ corresponding to $N=1$. For $N>1$, we will have a  $\mathbf{G}^\star$ matrix with size of $(2N+2)\times(2N+2)$. Its first and last entry on the main diagonal will still be $\e^{-\cj M\omega \tau}$ as in Eqs.~(\ref{eq:hmatrix_in_1M})-(\ref{eq:hmatrix_in_1M_even}). The middle part of $\mathbf{G}^\star$ will be filled with $N$ different $\xi_M$ in a similar way as in Eqs.~(\ref{eq:hmatrix_in_1M})-(\ref{eq:hmatrix_in_1M_even}), where the $n$-th $\xi_M$ corresponds to the $n$-th row in the PPIC. To intuitively understand this, notice that the horizontal relaxation actually confines the horizontal propagation of signals in the same arm. Thus, the light propagating in the first row will never go to the second row, which means the transfer functions of two different rows are decoupled.

An important implication from this example is that even under this simplifying horizontal relaxation, the final transfer function shown in Eq.~(\ref{eq:def_xim}), though it has an analytical form, does not provide a direct solution -- that is, does not provide us with a direct analytical filter synthesis method. This motivates our optimization-based synthesis method proposed in the next section.  

\section{Realization of Light Processing Functions}\label{sec:realization}

In this section, we explain how we utilize our derivation to efficiently synthesize light processing functions on an $N\times M$ square-mesh PPIC. Assume that we want to attain $N$ light processing functions represented by the complex transfer functions $\{U_n(\omega)|n=1,2\cdots,N\}$ specifying the magnitude and phase responses in a range $[\omega_{\text{min}},\omega_{\text{max}}]$. We choose $N_{\text{grid}}$ frequency points $\{\omega_1=\omega_{\text{min}},\omega_2=\omega_{\text{min}}+\Delta\omega,\cdots,\omega_{N_{\text{grid}}}=\omega_{\text{max}}\}$ in this desired angular frequency range with incremental step equal to $\Delta\omega$. Then we can define an error or cost function:
\begin{equation}\label{eq:cost_implicit}
    Cost=\sum_{k=1}^{N_{\text{grid}}}\sum_{n=1}^N \left|a_{2n,M}^{(I)}(\omega_k)-U_n(\omega_k)\right|^2
\end{equation}
Note that here we have made the dependence of $a_{2n,M}^{(I)}$ on the angular frequency explicit. If we can make the cost in Eq.~(\ref{eq:cost}) sufficiently small by adjusting all phase shifts $\{\theta,\phi\}$ of all vertical and horizontal TBUs, then we succeed to synthesize the $n$-th light processing functions at the bottom port of the $n$-th row (i.e., $\text{A}_{2n,M}$). This can be done by using an optimization technique: we minimize the cost in Eq.~(\ref{eq:cost}) with respect to all phase shifts  $\{\theta,\phi\}$:
\begin{equation}\label{eq:cost}
\rev{\min_{\mathbf{x}}\quad Cost=\sum_{k=1}^{N_{\text{grid}}}\sum_{n=1}^N \left|a_{2n,M}^{(I)}(\omega_k,\mathbf{x})-U_n(\omega_k)\right|^2}
\end{equation}
\rev{where we use $\mathbf{x}$ to collectively represent all phase shifts $\{\theta,\phi\}$ and make the dependence of $\mathbf{x}$ explicit in the cost function.}

However, the difficulty lies in the fact that this optimizatiothen problem is extremely high-dimensional. For an $N\times M$ square-mesh PPIC as shown in Fig.~\ref{fig:ppic}, it has $2(2N+1)M$ phase shifts in total. Considering a fairly small $10\times 10$ PPIC, there are already $420$ phase shifts to tune. To the best of our knowledge, such a high-dimensional optimization problem is inefficient to solve unless using a gradient descent method with analytical gradients. Specifically, non-gradient methods take a long time to converge, and gradient descent methods based on numerical differentiation require many function evaluations to calculate the gradient once. Importantly, in our case, we do have the analytical derivative $\partial{Cost}/\partial{\theta}$ or $\partial{Cost}/\partial{\phi}$ for any $\theta$ and $\phi$ based on our previous derivations, because the operations which relate $\theta$ (or $\phi$) to the variable $Cost$ are all differentiable. As a result, we can use gradient descent optimization to minimize Eq.~(\ref{eq:cost}) to perform the synthesis task. For details about how to calculate the gradient, please refer to Appendix \ref{sec:appendixb}.

We note that in some applications, the desired light processing functions only have requirements on the magnitude, but with no constraints on the phase. In such cases, we can choose $\{U_n(\omega)|n=1,2\cdots,N\}$ to be real functions representing the desired magnitude response and revise the cost in Eq.~(\ref{eq:cost}) as:
\begin{equation}\label{eq:cost_mag}
    Cost_{LinearMag}=\sum_{k=1}^{N_{\text{grid}}}\sum_{n=1}^N r_k \left||a_{2n,M}^{(I)}(\omega_k,\mathbf{x})|-U_n(\omega_k)\right|^2
\end{equation}
where $r_k$ ($k=1,2,\cdots,N_{\text{grid}}$) is a user-defined positive real scalar controlling the weight ratio. As will be demonstrated in our numerical results, we find that building upon Eq.~(\ref{eq:cost_mag}) and using logarithm magnitude works even better, especially for synthesising an optical filter where stop band and pass band have very different magnitude requirements. This logarithm cost is:
\begin{equation}\label{eq:cost_mag_log}
    Cost_{LogMag}=\sum_{k=1}^{N_{\text{grid}}}\sum_{n=1}^N r_k \left|\ln|a_{2n,M}^{(I)}(\omega_k,\mathbf{x})|-\ln U_n(\omega_k)\right|^2 \; .
\end{equation}

\section{Numerical Results}\label{sec:result}

In all our numerical experiments, we choose $n_{\text{eff}}=2.35$, $L=250$ um, $c=3\times10^8$ m/s, and $\alpha=0.99$. We do not take dispersion effects into account (i.e., $n_{\text{eff}}$ is considered to be constant and independent of $\omega$, which means that $n_{\text{g}}=n_{\text{eff}}=2.35$). Real waveguides do have dispersion, but this does not affect the method, as long as the dispersion of $n_{\text{eff}}$ can be described by an analytically derivable function (e.g., with the help of $n_\text{g}$). \rev{Moreover, we emphasize that in high refractive index contrast platforms, $n_{\text{eff}}$ usually depends on frequency and the dispersion effect causes a narrow free spectral range in the PPIC.} Before moving on, we define a value for later simplicity:
\begin{equation}\label{eq:delta_f}
    \Delta f = \frac{c}{n_{\text{g}}L}=\frac{3\times 10^8}{2.35\times 250\times10^{-6}}\approx 510.638\text{ Ghz}
\end{equation}
When plotting the figures of frequency response, we will normalize the frequency X-axis following the rule: 
\begin{equation}\label{eq:normalize_f}
    f_{\text{norm}}=\frac{2}{\Delta f}(f-f_{\text{center}})
\end{equation}
where $f_{\text{norm}}$ and $f$ represent the frequency value after and before normalization, respectively. Here $f_{\text{center}}$ represents the center frequency:
\begin{equation}\label{eq:center_freq}
    f_{\text{center}}=\frac{c}{\lambda_{\text{center}}}=\frac{3\times10^{8}}{1550\times10^{-6}}\approx 193.548 \text{ Thz} \; .
\end{equation}
For instance, Eq.~(\ref{eq:normalize_f}) will map $\{f_{\text{center}}-0.5\Delta f, f_{\text{center}},f_{\text{center}}+0.5\Delta f\}$ to $\{-1,0,1\}$, respectively. Recall that free spectral range (FSR) represents the periodicity of the frequency response when interference occurs. Provided our introduced notation $\Delta f$, these statements are equivalent: (i)~$\text{FSR}={c}/{n_{\text{g}}(KL)}=\Delta f / K$, and (ii)~in the normalized frequency figure, a range of $[-1,1]$ corresponds to $K$ periods, or one period has length $2/K$. Defining the value $\Delta f$ and plotting the frequency response on a normalized frequency axis give us a consistent way to visualize the results in different examples. 

Our algorithm is implemented in Python, \rev{and all our numerical experiments are performed on the same RedHat Linux server with 16 Intel Xeon E7-4850 CPUs working at 2.1GHz.} The initial guess required by the gradient descent optimization is randomly generated, consistent with our claim that our synthesis method does not require human design knowledge. However, we emphasize that in most of our examples, we are optimizing an interferometric system with many phase variables, and thus the cost function for most configurations will have many local peaks and valleys. The specific configuration coming out of the optimization algorithm will therefore depend strongly on the initial condition. Table~\ref{table:details_experiments} comprehensively lists the detailed information of all our experiments. In the following paragraphs, we comment on each case.

\begin{figure}[!htb]
    \centering
    \includegraphics[width=1.0\linewidth]{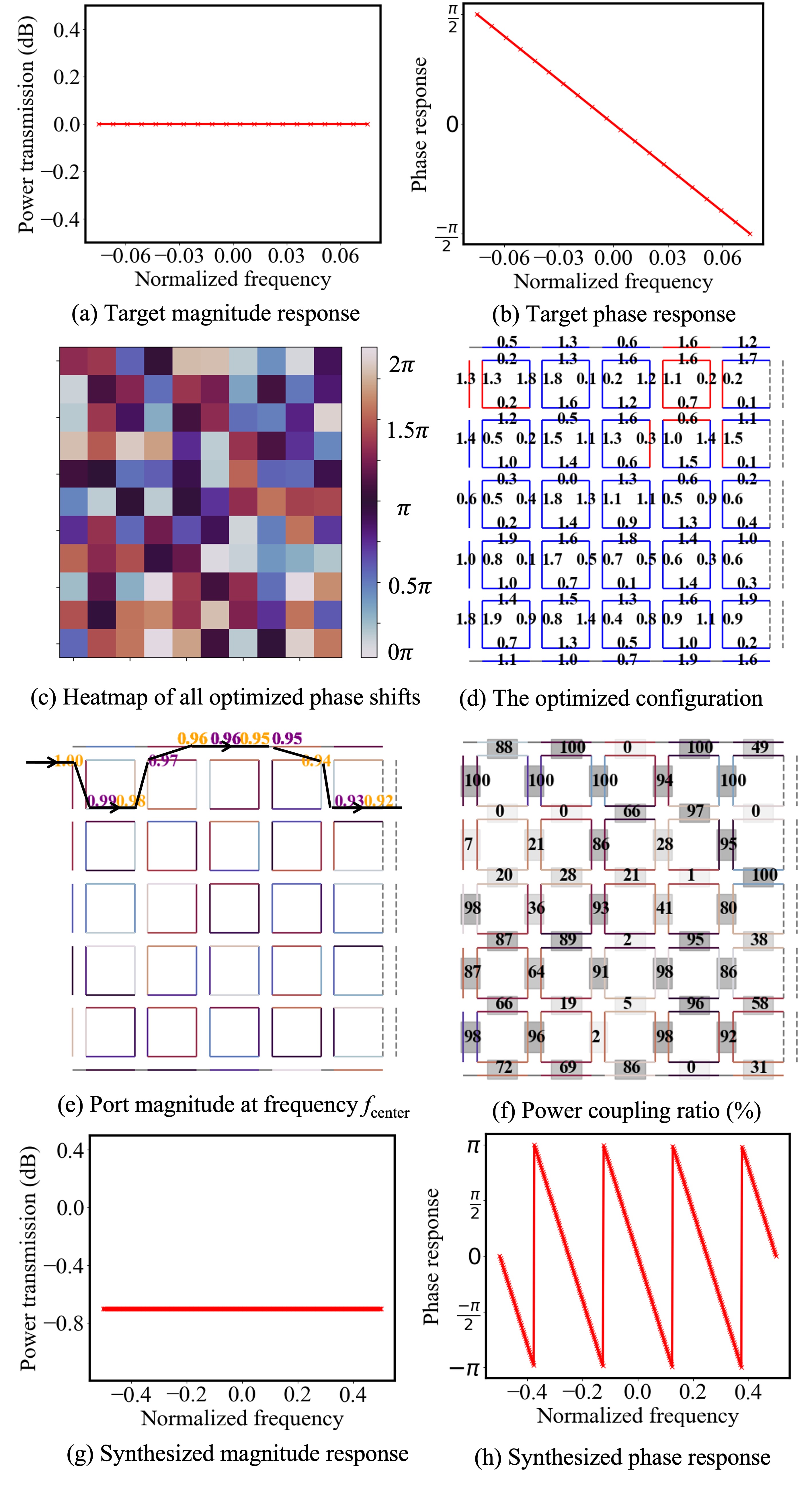}
    \caption{Case 1: routing. (a)~and (b)~show the target response $U(\omega)$ with magnitude normalized to input, and phase, respectively, used in the cost function. (c)~shows a heat map of all optimized phase shift values \secrev{(see Appendix \ref{sec:appendix_added} for colored cell ordering details)}. (d)~shows the resulting optimized configuration ($\pi$ omitted). Red lines are those phase shift changes larger than $0.2\pi$ before and after optimization, while blue lines are those with changes smaller than $0.2\pi$. (e)~shows the port magnitude, orange for inward direction and purple for outward direction (refer to Fig.~\ref{fig:ppic_name_convention} for definition). Port magnitudes less than 0.2 are not drawn. The light path is plotted in black. (f)~shows the power coupling ratio (i.e., $\cos^2\frac{\phi-\theta}{2}$) of each TBU with a percentage in a shaded bounding box. The edge color of the TBU shows common phase shift $\frac{\pi-\phi-\theta}{2}$; See Appendix \ref{sec:appendixc}. (g)~and (h)~show the frequency response that the optimized configuration is able to achieve. The square meshes in (e) and (f) share the same color-bar, shown in (c).}
    \label{fig:res_case1}
\end{figure}

For case 1, we consider routing the input light to an output port with minimum cost over the entire frequency band. Results are shown in Fig.~\ref{fig:res_case1}. The synthesized path shown in Fig.~\ref{fig:res_case1}~(e) has gone through eight TBUs. Thus, according to Eq.~(\ref{eq:our_compact_S}), we know that the synthesized configuration has a phase accumulation corresponding to $8L$, or more specifically, that the output port has a $\e^{-\cj\frac{\omega n_{\text{eff}}8L}{c}}$ dependence. This implies that we should witness a phase change of $2\pi$ over a frequency range of ${c}/{n_{\text{g}}(8L)}=\Delta f/8$, i.e., an interval with length $0.25$ in the normalized frequency figure. This is indeed the case as shown in Fig.~\ref{fig:res_case1}~(h). Also, if zooming in, Fig.~\ref{fig:res_case1}~(h) is exactly the same as Fig.~\ref{fig:res_case1}~(b). Since we have considered a loss term $\alpha=0.99$ in our compact TBU model, the synthesized normalized power transmission shown in Fig.~\ref{fig:res_case1}~(g) cannot reach 0~dB. We see that the synthesized light path shown in Fig.~\ref{fig:res_case1} relies on the top line and passes through eight TBUs, and a quick calculation shows $20\log0.99^8\approx -0.70$, consistent with Fig.~\ref{fig:res_case1}~(g). Last, but not least, Fig.~\ref{fig:res_case1}~(d) also demonstrates that only the first few rows have been adjusted by the optimization routine. This is as expected since our input and output port are both located at the top part of the mesh.

\begin{figure}[!htb]
    \centering
    \includegraphics[width=1.02\linewidth]{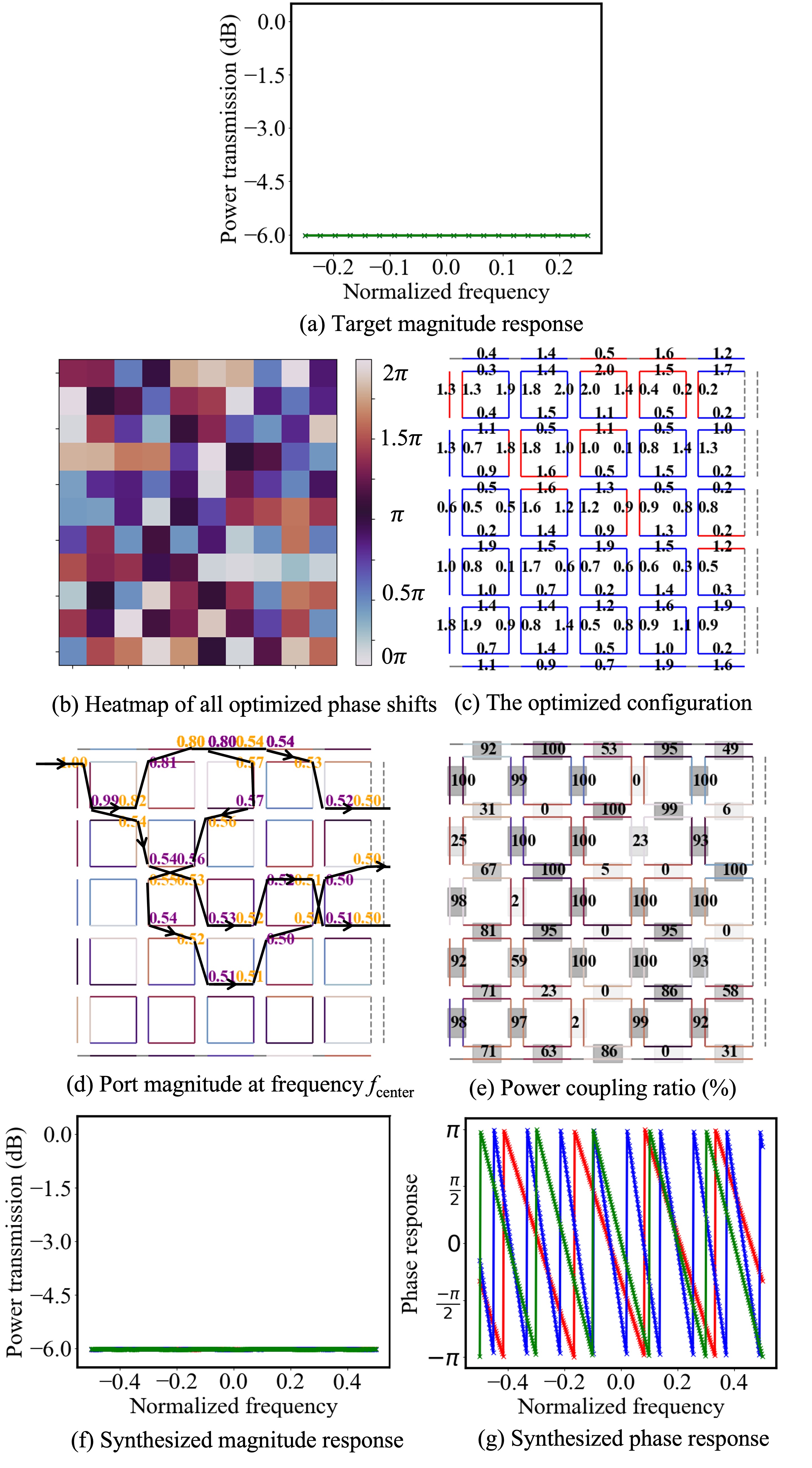}
    \caption{Case 2: splitting. (a)~target equal three-way split magnitude response (normalized to the input). (b)~heat map of all optimized phase shift values. (c)~optimized configuration ($\pi$ omitted). (d)~port magnitude. (e)~power coupling ratio and common phase shift. (f)~and (g) show the frequency response that the optimized configuration is able to achieve. There are three lines colored in red, blue, and green in (a), and they overlap here and in (f).}
    \label{fig:res_case2}
\end{figure}

For case 2, we consider equal power splitting to three output ports. Results are shown in Fig.~\ref{fig:res_case2}. We note that due to reciprocity, combining three light inputs can also be readily solved. As shown in Fig.~\ref{fig:res_case2}~(d), the three light paths pass through $8$, $10$, and $17$ TBUs, respectively, implying the three output responses should have phase accumulations corresponding to $8L$, $10L$, and $17L$. Namely, we will see a phase change of $2\pi$ over a frequency range of $\Delta f /8$, $\Delta f /10$, and $\Delta f /17$, respectively, corresponding to an interval with length $2/8$, $2/10$, and $2/17$ in the normalized frequency figure. As shown in Fig.~\ref{fig:res_case2}~(g), these correctly reflect the 4, 5, and 8.5 periods in the interval of $[-0.5,0.5]$. One subtlety here is that when designing the target function $U(\omega)$, we consider the power loss due to $\alpha=0.99$ and provide some margin in advance. Namely, the target magnitude chosen here is $0.5$ on a linear scale (i.e., about $-6.0$ dB in Fig.~\ref{fig:res_case2}~(a)), such that $ 0.5^2\times 3 =0.75<1.0$. Alternatively, choosing all three target magnitudes to be $\sqrt{1/3}$ on a linear scale would be problematic. From a numerical perspective, the optimization routine would seek to push all three output magnitudes to $\sqrt{1/3}$, but since this is unattainable simultaneously due to the loss term $\alpha$, it could happen that the resulting three outputs would be unequal (e.g., $\sqrt{0.30}, \, \sqrt{0.31}, \, \sqrt{1/3}$). Using a target magnitude which is attainable as we do here can prevent this issue. However, one side effect of a pre-provided power loss margin is that it might encourage the light path to go through more TBUs. For instance, the zigzag light path with $17L$ in this example is only one possible solution. It is obvious from Fig.~\ref{fig:res_case2}~(d) that this light path could propagate to the right bottom direction at the port with magnitude $0.57$ in the middle, instead of going to the left bottom as it currently does. 

\begin{figure}[!htb]
    \centering
    \includegraphics[width=1.0\linewidth]{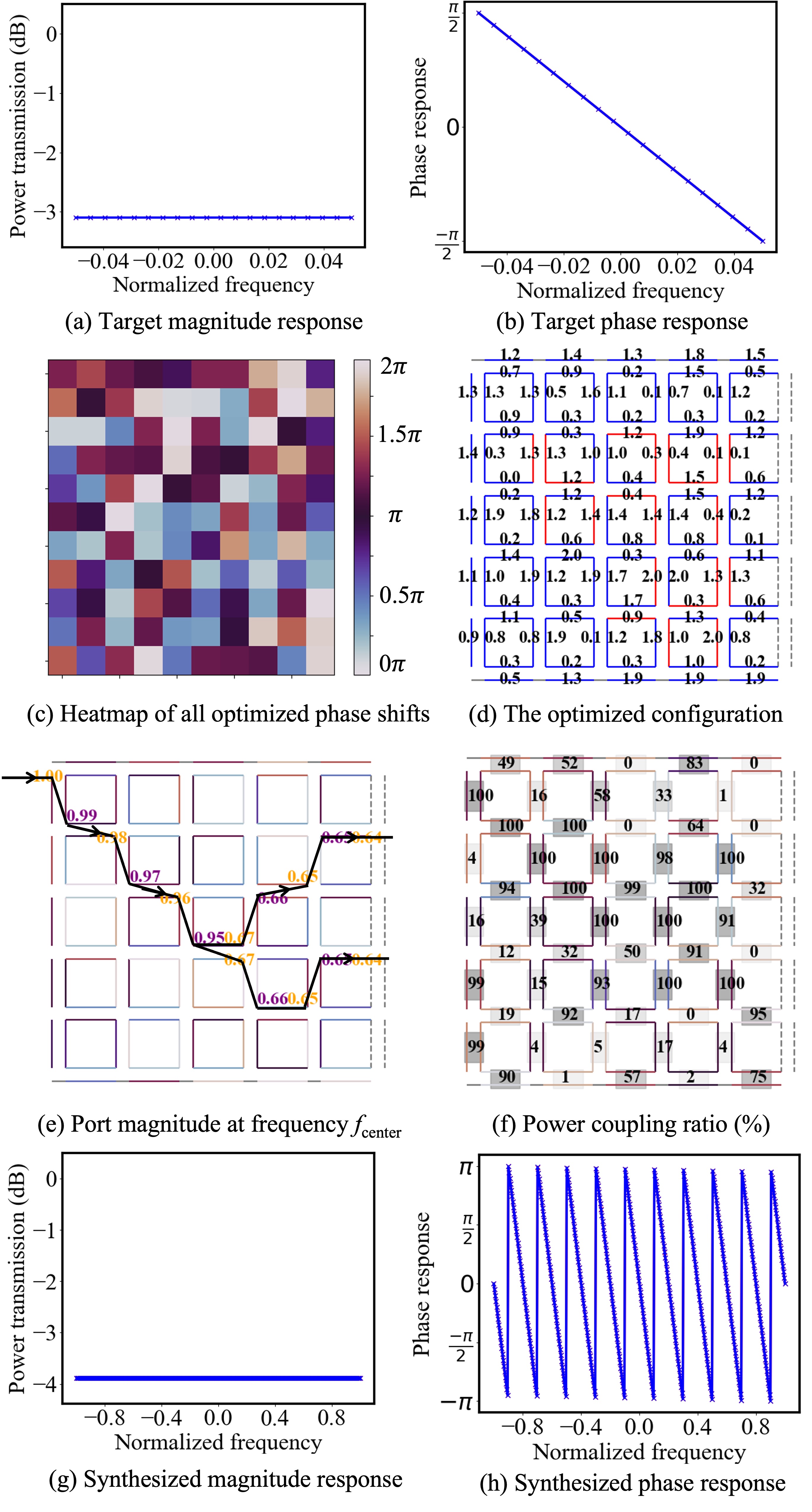}
    \caption{Case 3: coherent two-way splitting. (a)~and (b)~respectively show the target equal magnitude split with equal phase response. (c)~heat map of all optimized phase shift values. (d)~optimized configuration ($\pi$ omitted). (e)~port magnitude. (f)~power coupling ratio and common phase shift. (g)~and (h)~show the frequency response that the optimized configuration is able to achieve. There are two lines colored in red and blue in (a), and they overlap here and in (b), (g), and (h).}
    \label{fig:res_case3}
\end{figure}

For case 3, we consider coherent splitting. Namely, we want to split the input light to two output ports but now with identical phase. Results are shown in Fig.~\ref{fig:res_case3}. As seen in Fig.~\ref{fig:res_case3}~(e), both light paths pass through 10 TBUs, implying that a frequency range of $\Delta f/10$ (i.e., an interval with length $0.2$ in the normalized frequency figure) is required for a phase change of $2\pi$. This is also confirmed by Fig.~\ref{fig:res_case3}~(h). Moreover, we note that in a $5\times 5$ square mesh, without using the top or bottom line, the minimum number of TBUs required to propagate light from a port at left to a port at right is $10$. Moreover, the optimization routine obtains a synthesized result that seems natural and readily understandable. Namely, we chose the output port row indices to be $3$ and $7$ in this case, while the input port row index was $1$ (see Table~\ref{table:details_experiments}). The resulting synthesized light path first goes from top left to the bottom right direction without any splitting, and then approximately stops at the middle between the output ports. Then it performs a 50\%:50\% power splitting and the resulting two light paths keep propagating without further splitting all the way to the output ports. This approach of first propagating to the middle followed by a 50\%:50\% splitting is a generic strategy to synthesize one-input to two-output coherent splitting, and is automatically found by the optimization.


\begin{figure}[!htb]
    \centering
    \includegraphics[width=1.0\linewidth]{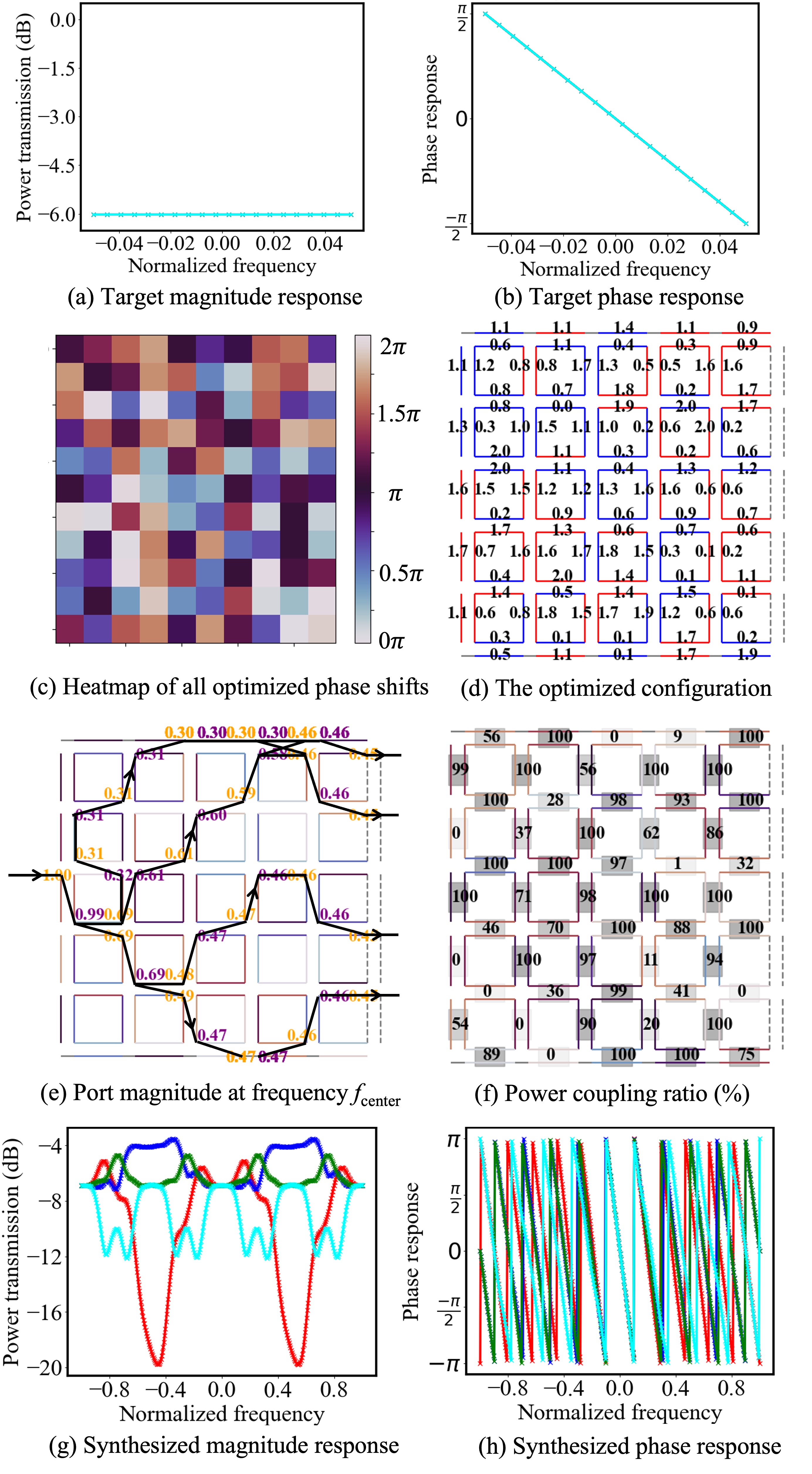}
    \caption{Case 4: coherent four-way splitting. (a)~and (b)~respectively show the target equal four-way magnitude split and phase response. (c)~heat map of all optimized phase shift values. (d)~optimized configuration ($\pi$ omitted). (e)~port magnitude. (f)~power coupling ratio and common phase shift. (g)~and (h)~show the frequency response that the optimized configuration is able to achieve. There are four lines colored in red, blue, green, and cyan in (a), and they overlap here and in (b).}
    \label{fig:res_case4}
\end{figure}

For case 4, we consider a more complicated version of case 3. Now, we attempt to do coherent splitting to four output ports. Results are shown in Fig.~\ref{fig:res_case4}. Due to the structure of the square mesh, it is actually impossible to find four light paths all with the same length, meaning that the goal in this case is unachievable. Specifically, because the four output ports do not belong to the same clockwise/counter-clockwise sub-mesh, there will be at least one $L$ light path difference. As shown in Fig.~\ref{fig:res_case4}~(e), it seems that the optimization attempts to utilize interference to approach this unattainable goal as close as possible. From Fig.~\ref{fig:res_case4}~(g), we see that the red and blue curves both have an FSR equal to $0.5\Delta f$, while the green and cyan curves both have an FSR equal to $0.25\Delta f$. The difference of FSRs also indicates that we cannot achieve coherent splitting at an arbitrary frequency point, since these paths have periodicity mismatch. This is also verified in Fig.~\ref{fig:res_case4}~(g) and (h). Our synthesized results do satisfy the given targets shown in (a)~and (b): the optimization achieves coherent splitting in the normalized range $[-0.05,0.05]$, which corresponds to around a 25 Ghz range in reality. However, we also notice that outside this range, the optimization cannot always achieve coherent splitting. An important note is that there are several rings in the synthesized configuration in this case, and explains why we obtain a frequency-dependent response in Fig.~\ref{fig:res_case4}~(g). However, port magnitudes associated with some of the rings are smaller than 0.2, and thus are not drawn. 

\begin{figure}[!htb]
    \centering
    \includegraphics[width=1.0\linewidth]{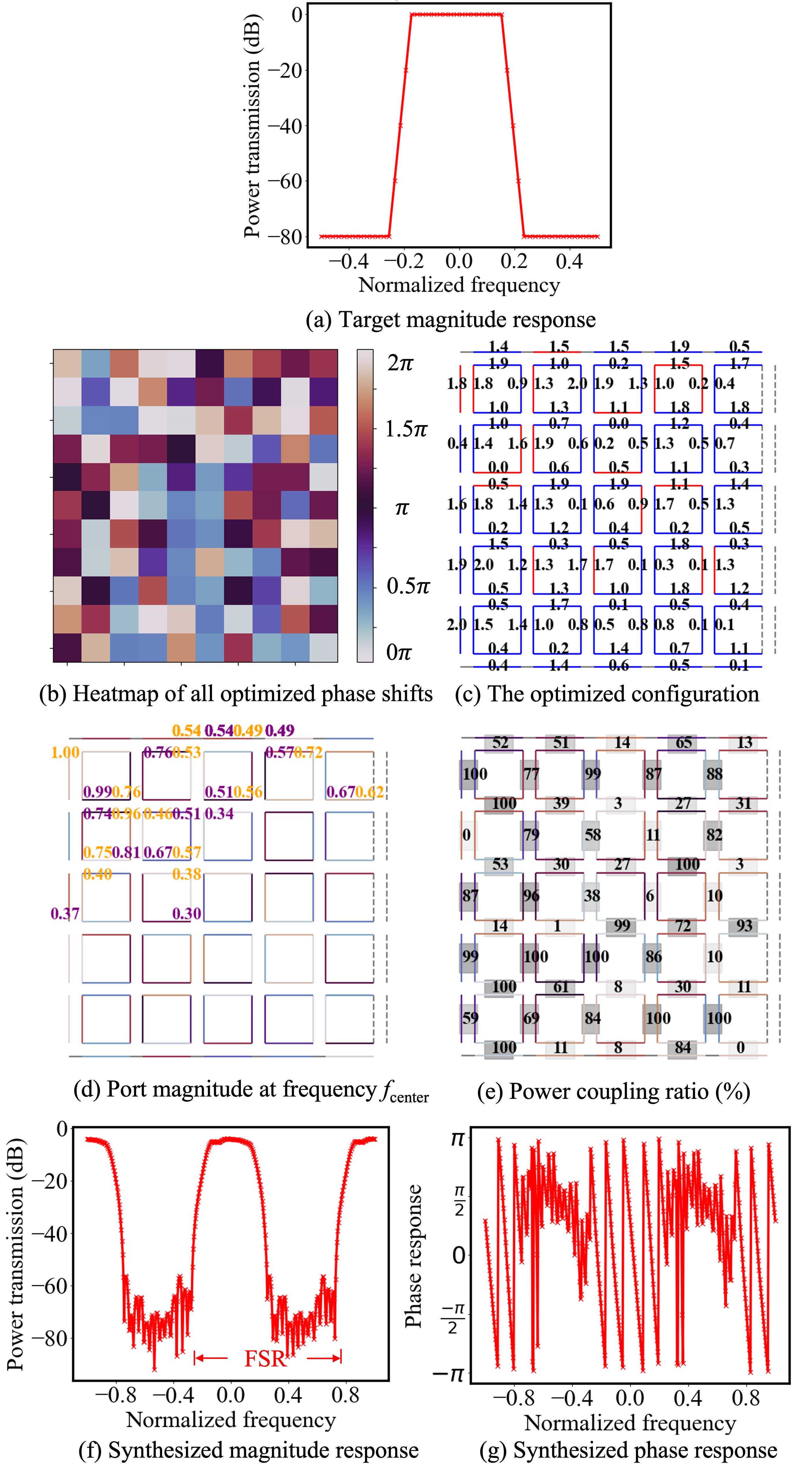}
    \caption{Case 5: optical filtering. (a)~target magnitude response. (b)~heat map of all optimized phase shift values. (c)~optimized configuration ($\pi$ omitted). (d)~port magnitude. (e)~power coupling ratio and common phase shift. (f)~and (g) show the frequency response that the optimized configuration is able to achieve. Note that in this case, only port magnitudes over 0.3 are plotted in (d) for clarity.}
    \label{fig:res_case5}
\end{figure}

For case 5, we consider optical filtering. Results are shown in Fig.~\ref{fig:res_case5}. As seen in Fig.~\ref{fig:res_case5}~(d), many rings have formed in the obtained configuration. We successfully achieve near 0~dB in the pass band, and about $-70$~dB in the stop band. The FSR is about $0.5\Delta f$ as depicted in Fig.~\ref{fig:res_case5}~(f). 

\begin{figure}[!htb]
    \centering
    \includegraphics[width=1.0\linewidth]{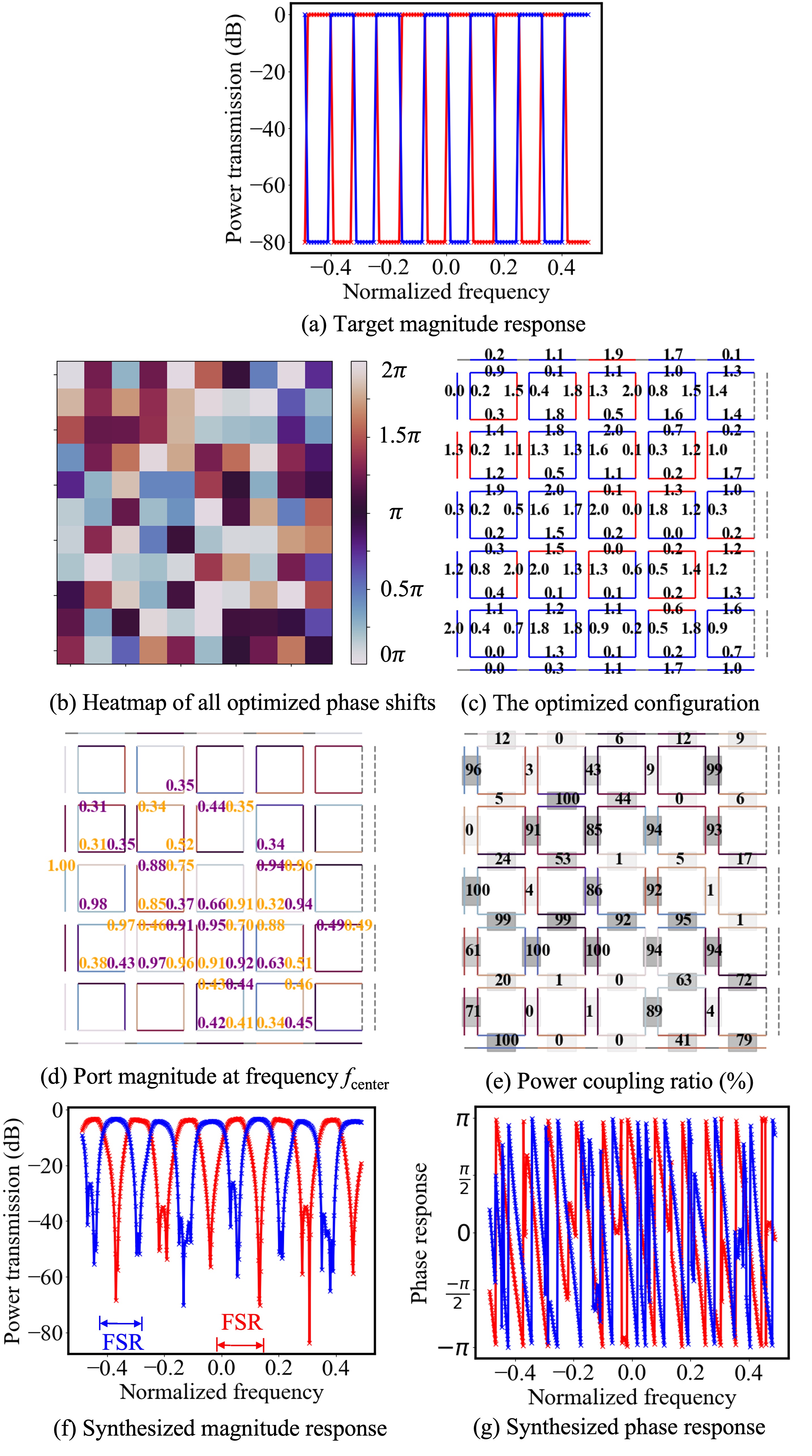}
    \caption{Case 6: wavelength division multiplexing. (a)~target magnitude response. (b)~heat map of all optimized phase shift values. (c)~optimized configuration ($\pi$ omitted). (d)~port magnitude. (e)~power coupling ratio and common phase shift. (f)~and (g) show the frequency response that the optimized configuration is able to achieve. Note that in this case, only port magnitudes over 0.3 are plotted in (d) for clarity.}
    \label{fig:res_case6}
\end{figure}

As case 6, we consider two-way wavelength division multiplexing (WDM), also called an optical interleaver, where the spectrum is separated into even and odd frequency channels over two outputs. From the results in Fig.~\ref{fig:res_case6}~(d), it is clear that many rings have formed in the optimized configuration. Moreover, (d) is plotted at the central frequency, and thus the other output port magnitudes are less than 0.3 and not drawn. 

\begin{figure}[!htb]
    \centering
    \includegraphics[width=1.0\linewidth]{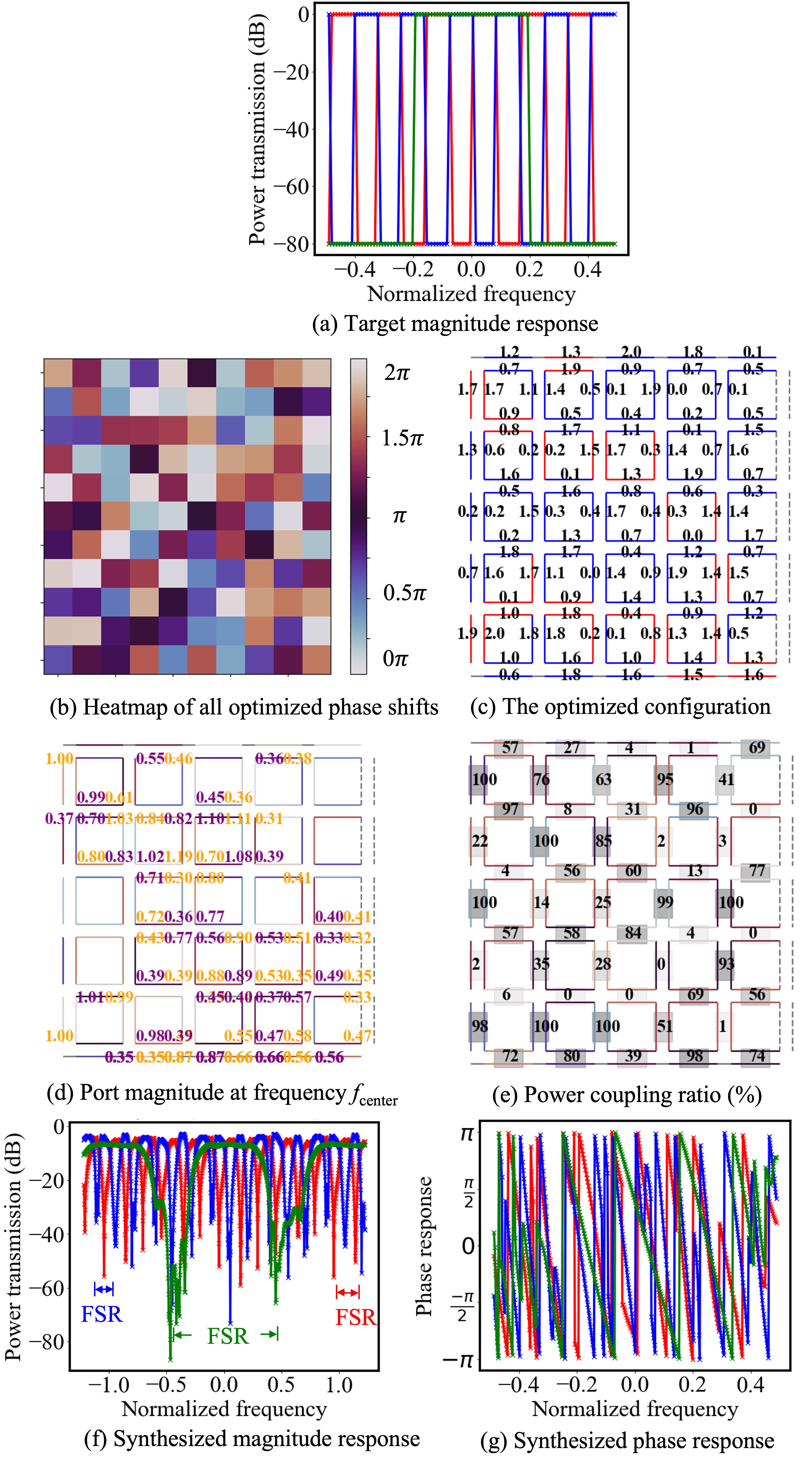}
    \caption{Case 7: Simultaneously synthesize two light processing functions for two in-phase inputs. Figure caption is similar to that of Fig.~\ref{fig:res_case1}, except that in this case, only port magnitudes over 0.3 are plotted in (d) for clarity.}
    \label{fig:res_case7}
\end{figure}

For case 7, we consider synthesizing two light processing functions (WDM and optical filtering) at the same time, given two in-phase inputs. Namely, we provide a complex input $1.0+0.0\cj$ at $A_{1,0}$ and a complex input $0.0+1.0\cj$ at $A_{10,0}$. The two output ports for WDM are $A_{2,5}$ and $A_{6,5}$, while that for filtering is $A_{10,5}$. Results are shown in Fig.~\ref{fig:res_case7}. Note that in Fig.~\ref{fig:res_case7}~(d), we see that some inner port magnitudes are larger than $1.0$. This is possible because (i)~the total input power is $2.0$, and (ii)~when a ring is formed, it can lead to the 'intensity build-up' phenomenon~\cite{bogaerts2012silicon,heebner2008optical} near resonance.

\rev{To better quantify the performance of our method, we implement two baseline methods for comparison: (i)~differential evolution, a population-based gradient-free global optimization approach; and (ii)~gradient descent optimization with numerical differentiation. Table~\ref{tab:runtime_comparison} summarizes the run-time of our method and the two baselines. We see that our proposed method achieves about \improve x computation time cost reduction compared with the implemented baseline methods.}

\begin{table*}[!thb]
\small
\centering

\begin{tabular}{cccccccc}
\toprule & Input Port &  Output Port  &  Target(s) & Cost & Results & Run-time & Phase Acc / FSR * \\
\midrule
No. 1: Routing   & $A_{1,0}$ & $A_{2,5}$  & mag, phase   & Eq.~(\ref{eq:cost}) & Fig.~\ref{fig:res_case1} & $0.27$ mins & $8L$\\
No. 2: Splitting   & $A_{1,0}$ & $A_{2,5},\,A_{4,5},\,A_{6,5}$  & mag   & Eq.~(\ref{eq:cost_mag}) & Fig.~\ref{fig:res_case2} & $1.09$ mins & $8L,\,10L,\,17L$\\
No. 3: Splitting (c)~  & $A_{1,0}$ & $A_{3,5},\,A_{7,5}$  & mag, phase   & Eq.~(\ref{eq:cost}) & Fig.~\ref{fig:res_case3} & $0.63$ mins & $10L,\,10L$\\
No. 4: Splitting (c)~  & $A_{5,0}$ & $A_{1,5},\,A_{3,5},\,A_{7,5},\,A_{9,5}$  & mag, phase   & Eq.~(\ref{eq:cost}) & Fig.~\ref{fig:res_case4} & $4.48$ mins & $\frac{\Delta f}{2},\,\frac{\Delta f}{2},\,\frac{\Delta f}{4},\,\frac{\Delta f}{4}$\\
No. 5: Filtering   & $A_{1,0}$ & $A_{2,5}$  & mag   & Eq.~(\ref{eq:cost_mag_log}) & Fig.~\ref{fig:res_case5} & $81.59$ mins & $\frac{\Delta f}{2}$\\
No. 6: WDM   & $A_{5,0}$ & $A_{3,5},\,A_{7,5}$  & mag   & Eq.~(\ref{eq:cost_mag_log}) & Fig.~\ref{fig:res_case6} & $108.25$ mins & $\frac{\Delta f}{12},\, \frac{\Delta f}{12}$\\
No. 7: WDM \& Filtering   & $1\,@\,A_{1,0},\,1\cj\,@\,A_{10,0}$ & $\{A_{2,5},\,A_{6,5}\},A_{10,5}$  & mag   & Eq.~(\ref{eq:cost_mag_log}) & Fig.~\ref{fig:res_case7} & $110.34$ mins & $\frac{\Delta f}{12},\, \frac{\Delta f}{12},\,\frac{\Delta f}{2}$ \\
\bottomrule
\end{tabular}
\caption{Detailed information for all our experiments. All are performed on a $5\times 5$ square mesh. `(C)' is short for `coherent'. `Mag' is short for `magnitude'. When using the logarithm cost Eq.~(\ref{eq:cost_mag_log}), we set $r_k$ to $10$ and $1$ for frequency points in the pass band and stop band, respectively. If there is no stop band (e.g., case 2), $r_k$ is set to $1.0$ for all $k$. * In case 1, 2, and 3, the synthesized results have no interference, we use phase accumulation to depict how many TBUs the light path passes through (e.g., $8L$). In case 4, 5, 6, and 7, interference occurs and it becomes less clear that the light path goes through a specific number of TBUs. Thus, we use the metric FSR (e.g., $\Delta f/2$) in these cases.}

\label{table:details_experiments}
\end{table*}

\begin{table}[!htb]
    \centering
    \begin{tabular}{cccc}
    \toprule
                             & Ours    & DE   & ND \\
     No. 1: Routing          & $0.27$               & $>100$    & $\approx 22.44$ \\
     No. 2: Splitting        & $1.09$               & $>100$    & $\approx 78.02$ \\
     No. 3: Splitting (c)~   & $0.63$               & $>100$    & $\approx 61.72$ \\
     No. 4: Splitting (c)~   & $4.48$               & $>100$    & $\approx 80.26$ \\
     No. 5: Filtering        & $81.59$              & $>400$    & $>400$ \\
     No. 6: WDM              & $108.25$             & $>400$    & $>400$ \\
     No. 7: WDM \& Filtering & $110.34$             & $>400$    & $>400$ \\
    \bottomrule
    \end{tabular}
    \caption{\rev{Run-time (minutes) comparison of our method with DE and ND. DE is short for differential evolution, a population-based gradient-free global optimization approach. ND is short for gradient descent optimization with numerical differentiation. We stop DE/ND when the synthesized results attain similar cost values as our method or similar curve shapes in the magnitude or/and phase response figures. The `$>$' sign indicates that the corresponding algorithm's result is not comparable to ours within the specified time.}}
    \label{tab:runtime_comparison}
\end{table}

\begin{figure}[!htb]
    \centering
    \includegraphics[width=0.95\linewidth]{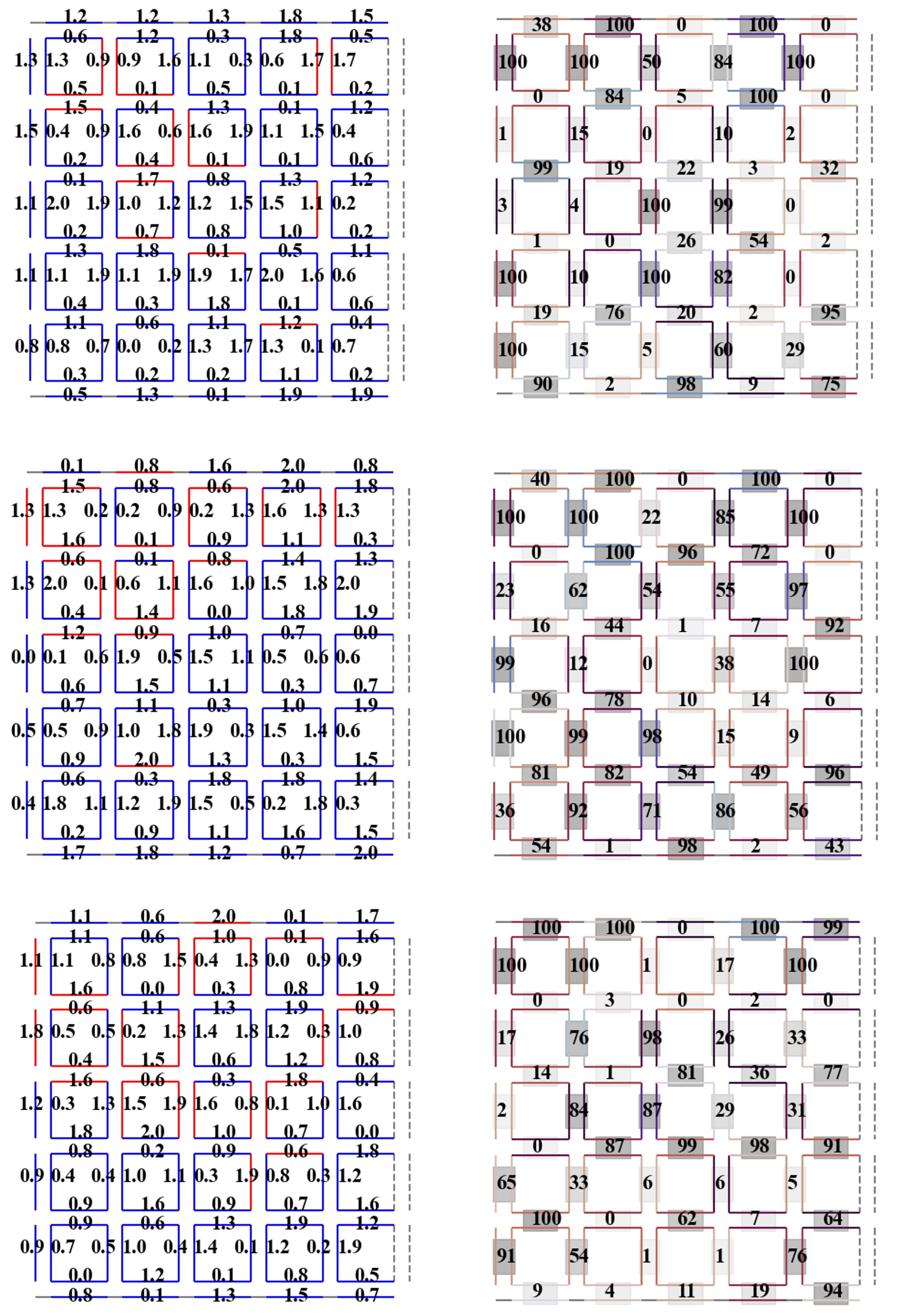}
    \caption{\rev{Three different configurations are obtained under three random initializations; all satisfy the goal of routing in Case 1. Each row represents one synthesized configuration. Fig.~\ref{fig:res_case1} is not included here. Left column: \secrev{The optimized configuration ($\pi$ omitted)}. Right column: Power coupling ratio (\%) and common phase shift. \secrev{All synthesized magnitude responses are identical to those in Fig.~\ref{fig:res_case1} (g), hence not shown.}} }
    \label{fig:case1_random}
\end{figure}

\begin{figure}[!htb]
    \centering
    \includegraphics[width=1.0\linewidth]{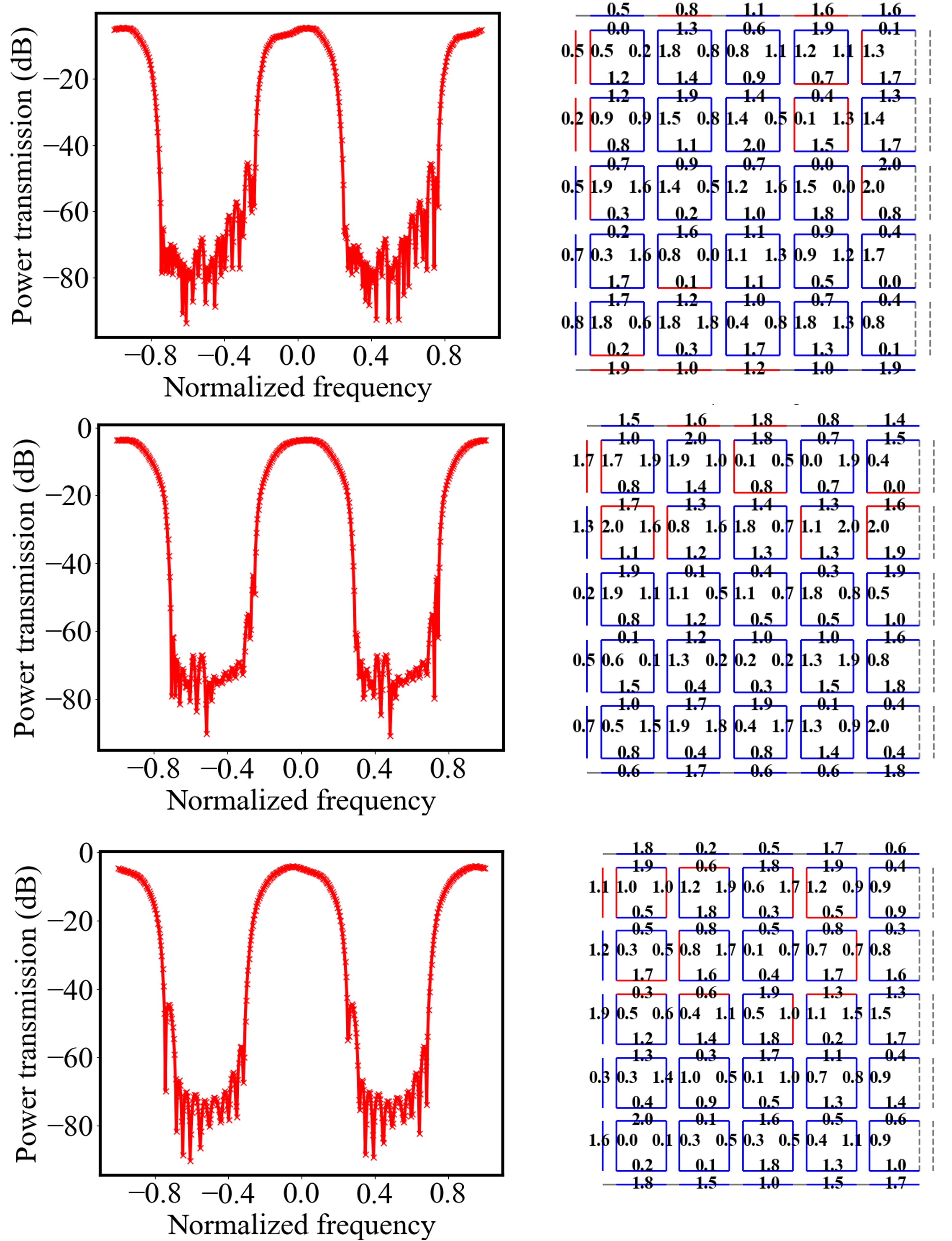}
    \caption{\rev{Three different configurations are obtained under three random initializations; all satisfy the goal of filtering in Case 5. Each row represents one synthesized configuration. Fig.~\ref{fig:res_case5} is not included here. Left column: Synthesized magnitude response. Right column: \secrev{The optimized configuration ($\pi$ omitted).}} }
    \label{fig:case5_random}
\end{figure}

\rev{We emphasize that gradient descent optimization (with potentially non-convex cost functions such as ours) is known to be only able to find local minima, and thus the specific configuration coming out of the optimization algorithm will depend strongly on the initial condition. To justify the practical utility of the proposed method, we also need to show that even with different initializations, the optimization routine can always yield a good result. Due to space limitations, we take Cases 1 and 5 as examples. We run our method with different initializations and plot the results in Fig.~\ref{fig:case1_random} and Fig.~\ref{fig:case5_random}. These demonstrate the robustness of our method to random initialization. Note that for our applications, we do not necessarily need a global optimum, while a locally optimal configuration is already sufficient. Note that when the PPIC size further scales up, we would expect the optimization result to be more strongly impacted by the initialization, because more local optima might exist for a higher dimensional optimization problem.}

\section{Conclusions, Limitations, and Future Work}\label{sec:discussion}

In this paper, we propose an efficient synthesis method that can be applied to realize configurations for a wide range of light processing functions on a square-mesh PPIC. The key property that makes our method efficient is that we  analytically derive the gradients of the mean squared error, or the log ratio, between target and realized circuit response with respect to the tunable phase shifts based on scattering matrix theory. Then, a gradient descent optimization can be carried out to synthesize the desired light processing functions at time scales of minutes.

\paragraph{Other PPIC connection topologies.} We consider a square mesh in this paper because it provides the clearest derivation of the scattering matrix elements, compared to triangular and hexagonal meshes, due to the fact that the TBUs are placed either vertically or horizontally. 
Nevertheless, we emphasize that even though identifying column $1, 2,\cdots, M$ as shown Fig.~\ref{fig:ppic} is harder for a triangular or hexagonal mesh, it is still possible, and thus our method is also applicable to these topologies. \secrev{For example, the authors in~\cite{perez2019scalable} have successfully derived a system-level transfer function for a hexagonal mesh using a similar approach as ours.} However, when a mixture of triangular, square, and hexagonal mesh is used, or an arbitrary connection of TBUs is adopted, the current implementation of our method \secrev{or~\cite{perez2019scalable}} can fail because it may not be possible to divide the circuit into columns, and \secrev{we both build} the scattering matrix iteratively going through column by column. In the future, we will expand our approach to arbitrary connection topologies.

\paragraph{Dealing with non-ideality.} Two assumptions used in this paper (i.e., omitting the last column and assuming ideal yellow lines in Fig.~\ref{fig:ppic}) are only for ease of mathematical notation. These assumptions can be relaxed, consistent with our method. In a real world scenario, dispersion effects can exist either due to the frequency-dependent effective index of the waveguide, or due to a frequency-dependent power coupling ratio in the 50\%:50\% directional couplers. Furthermore, each building block TBU might be slightly different due to process variations in manufacturing. All of these non-idealities can be addressed as long as we can describe them as differentiable functions of frequency. Indeed we can do so, for example by using a Taylor expansion and introducing a group index for dispersion.

\rev{Another major source of non-ideality comes from the thermal cross-talk of heaters, which we have not considered in the main text. However, by using thermal eignemode decomposition~\cite{Maziyar2019canceling}, our proposed method remains applicable under a change of optimized variables and can account for thermal cross-talk. A similar treatment can be adopted for other non-idealities of actuators in PPICs. Please refer to Appendix~\ref{sec:appendixd} for details.}

\rev{Last, but not least, we emphasize that when non-idealities (e.g., process variation, dispersion effect, or beamsplitting error) are considered, a more complex variant of the proposed compact model might be needed. Please refer to Appendix~\ref{sec:appendixe} for details.}


\paragraph{Numerical considerations. }
As shown in our numerical results, choosing an appropriate cost function is of crucial importance, especially in a case when both stop band and pass band are present at the same time. In our paper, we use a weighted logarithm cost function for such cases, but note that other options also exist, such as a combination of linear and logarithm cost~\cite{wang2022programmable}. The choice of cost function can substantially impact the optimized results, and it will be interesting to consider if better cost functions exist. When doing so, it will be beneficial to explore among differentiable cost functions, since gradient descent optimization is preferred in this problem. Exploration could also overcome another limitation of our current cost functions. Ideally, we want the cost to be zero if the achieved rejection ratio (e.g., $|-80|$ dB) is already larger than the target (e.g., $|-70|$ dB) in the stop band. However, implementing this threshold strategy might degrade the performance of gradient descent optimization as it will introduce non-differentiable points in the optimization search space.

We also note that the matrix $\mathbf{V}$ shown in Eq.~(\ref{eq:transformed_vertical_transferfunc}) is not invertible when the two phase shifts $\{\theta,\phi\}$ have a $\pi$ difference (i.e., $\theta=\pi+\phi$ or vice versa). In this case, our approach will fail. This is consistent with our intuition: when the phase shifts have a $\pi$ difference, the vertical TBU is in the bar state, and knowing all port magnitudes related to `A' does not confer any knowledge on the the port magnitudes related to `B' (see Fig.~\ref{fig:ppic_name_convention} and Eq.~(\ref{eq:transformed_vertical_transferfunc})). In a real numerical implementation, this means that if the phase difference $|\phi-\theta|$ is close to $\pi$, then the associated $\mathbf{V}$ will be ill-conditioned, and our simulated frequency response at ports and the gradients might be inaccurate. Fortunately, in our optimized results, we have not encountered this issue. Observant readers might find that, for example, in Fig.~\ref{fig:res_case6}~(e), there exist a few vertical TBUs with power coupling ratio reported as $0$, implying that they are in bar states. However, this is because we only display up to integer percentage when drawing the figure for space reasons. To support the claim made in the main text, we have printed the condition number of $\mathbf{T}^\star$ defined in Eq.~(\ref{eq:tstar}) as a way to examine numerical stability when running the program. But we warn of the need to pay attention to this case. In the future, this problem should readily be solved when we expand our approach to support any connections, since at that time, our scattering matrix will be set up based on graph theory, without requiring inverting $\mathbf{V}$.


\vspace{6pt}
\paragraph{Funding.} Xiangfeng Chen and Wim Bogaerts received funding from the European Research Council through grant 725555 (PhotonicSWARM).

\paragraph{Disclosures.} The authors declare that there are no conflicts of interest related to this article.

\paragraph{Data availability.} Data underlying the results presented in this paper are not publicly available at this time but may be obtained from the authors upon reasonable request.

\bibliography{sample}

\appendix
\section*{Appendix}

\subsection{Justification for the Compact Model}\label{sec:appendixa}

Here we justify the validity of the compact TBU model given in Eq.~(\ref{eq:form_of_Smatrix}). Considering all possible locations of waveguides in a TBU, the most general form of the compact model is:
\begin{equation}\label{eq:form_of_Smatrix_general_wgs}
    \mathbf{F} = \frac{1}{2}\boldsymbol{\Lambda}_1\left[
    \begin{array}{cc}
        1 &  -\cj \\
        -\cj &  1
    \end{array}
    \right]
    \boldsymbol{\Lambda}_2
    \left[
    \begin{array}{cc}
        \e^{-\cj\theta} &  0 \\
        0 &  \e^{-\cj\phi}
    \end{array}
    \right]
    \boldsymbol{\Lambda}_3
    \left[
    \begin{array}{cc}
        1 &  -\cj \\
        -\cj &  1
    \end{array}
    \right]
    \boldsymbol{\Lambda}_4
\end{equation}
where we have defined $\boldsymbol{\Lambda}_i$ ($i=1,2,3,4$):
\begin{equation}
    \boldsymbol{\Lambda}_i = \left[
    \begin{array}{cc}
        \alpha_i\e^{-\cj\omega \frac{n_{\text{eff}} L_i}{c}} &  0 \\
        0 &  \alpha_i^\prime\e^{-\cj\omega \frac{n_{\text{eff}} L_i^{\prime}}{c}}
    \end{array}
    \right]
\end{equation}
and $\{L_i,L_i^\prime\}$ represent the length of waveguide in the upper and lower arm, respectively. For instance, here $\boldsymbol{\Lambda}_2$ represents the waveguides between the right DC and the middle phase shifts (PSs). As long as the waveguides are balanced in the upper and lower arm (i.e., $L_i=L_i^\prime$ and $\alpha_i=\alpha_i^\prime$ for all $i=1,2,3,4$), then Eq. (\ref{eq:form_of_Smatrix_general_wgs}) can be simplified as:
\begin{equation}\label{eq:form_of_Smatrix_wgs}
    \mathbf{F} = \frac{1}{2}\left[
    \begin{array}{cc}
        1 &  -\cj \\
        -\cj &  1
    \end{array}
    \right]
    \left[
    \begin{array}{cc}
        \e^{-\cj\theta} &  0 \\
        0 &  \e^{-\cj\phi}
    \end{array}
    \right]
    \left[
    \begin{array}{cc}
        1 &  -\cj \\
        -\cj &  1
    \end{array}
    \right]\alpha
    \e^{-\cj\omega \frac{n_{\text{eff}}L}{c}}
\end{equation}
where $L$ equals the sum of all $L_i$, and $\alpha$ equals the product of all $\alpha_i$:
\begin{equation}
    L=\sum_{i=1}^4 L_i,\quad \alpha=\prod_{i=1}^4 \alpha_i \; .
\end{equation}
Fortunately, the assumption of balanced waveguides is only a mild one: imbalanced waveguides are seldom used in PPICs, so our compact model remains applicable.

\subsection{Example of Calculating the Gradient}\label{sec:appendixb}

According to Eq.~(\ref{eq:only_consider_forward_output}) of the main text, we can calculate the derivative of $\mathbf{a}_{M}^{(I)}$ with respect to one phase shift $\theta$:
\begin{equation}\label{eq:deri_forward_output}
    \begin{aligned}
            \frac{\partial\mathbf{a}_{M}^{(I)}}{\partial \theta}&=\frac{\partial}{\partial \theta}(\mathbf{G}^\star\mathbf{a}_{0}^{(I)})=\frac{\partial\mathbf{G}^\star}{\partial \theta}\mathbf{a}_{0}^{(I)}
            =\frac{\partial (\mathbf{T}_{11}^\star-\mathbf{T}_{12}^\star\mathbf{T}_{22}^{\star,-1}\mathbf{T}_{21})}{\partial \theta}\mathbf{a}_{0}^{(I)}\\
            &=(\frac{\partial\mathbf{T}_{11}^\star}{\partial\theta}-\frac{\partial \mathbf{T}_{12}^\star}{\partial\theta}\mathbf{T}_{22}^{\star,-1}\mathbf{T}_{21}\\
            &\quad-\mathbf{T}_{12}^\star\frac{\partial \mathbf{T}_{22}^{\star,-1}}{\partial\theta}\mathbf{T}_{21}-\mathbf{T}_{12}^\star\mathbf{T}_{22}^{\star,-1}\frac{\partial\mathbf{T}_{21}}{\partial\theta})\,\mathbf{a}_{0}^{(I)}\\
            &=(\frac{\partial\mathbf{T}_{11}^\star}{\partial\theta}-\frac{\partial \mathbf{T}_{12}^\star}{\partial\theta}\mathbf{T}_{22}^{\star,-1}\mathbf{T}_{21}\\
            &\quad+\mathbf{T}_{12}^\star\mathbf{T}_{22}^{\star,-1}\frac{\partial \mathbf{T}_{22}^{\star}}{\partial\theta}\mathbf{T}_{22}^{\star,-1}\mathbf{T}_{21}-\mathbf{T}_{12}^\star\mathbf{T}_{22}^{\star,-1}\frac{\partial\mathbf{T}_{21}}{\partial\theta})\,\mathbf{a}_{0}^{(I)}\\
    \end{aligned}
\end{equation}
where in the last line we have used the property of derivative of matrix inverse. Namely, for a matrix $\mathbf{R}$, we have:
\begin{equation}
    \frac{\partial\mathbf{R}^{-1}}{\partial \theta}=-\mathbf{R}^{-1}\frac{\partial\mathbf{R}}{\partial\theta}\mathbf{R}^{-1} \; .
\end{equation}
Due to the block matrix definition:
\begin{equation}
    \mathbf{T}^\star=\left[\begin{array}{cc}
    \mathbf{T}^\star_{11} & \mathbf{T}^\star_{12} \\
     \mathbf{T}^\star_{21} & \mathbf{T}^\star_{22}
\end{array}
\right]
\end{equation}
we also have:
\begin{equation}\label{eq:block_deri_tstar}
        \frac{\partial\mathbf{T}^\star}{\partial\theta}=\left[\begin{array}{cc}
    \frac{\partial\mathbf{T}^\star_{11}}{\partial\theta} & \frac{\partial\mathbf{T}^\star_{12}}{\partial\theta} \\
     \frac{\partial\mathbf{T}^\star_{21}}{\partial\theta} & \frac{\partial\mathbf{T}^\star_{22}}{\partial\theta}
\end{array}
\right] \; .
\end{equation}
In  other words, if we know $\mathbf{T}^\star$ and $\frac{\partial\mathbf{T}^\star}{\partial\theta}$, then the derivative $\frac{\partial\mathbf{a}_{M}^{(I)}}{\partial \theta}$ is known. $\mathbf{T}^\star$ is straightforward as depicted in Eq.~(\ref{eq:tstar}).


As an example, if the $\theta$ we consider corresponds to the vertical TBU at Row 1 and Column 0, then only $\mathbf{T}^0$ depends on this $\theta$, so that we have:
\begin{equation}\label{eq:deri_tstar}
    \frac{\partial\mathbf{T}^\star}{\partial\theta}=\mathbf{P}^T\mathbf{T}^{M-1}\cdots\mathbf{T}^1\frac{\partial\mathbf{T}^0}{\partial\theta}\mathbf{P} \; .
\end{equation}
According to Eq.~(\ref{eq:expression_j_layer_transfer}) in the main text, we know:
\begin{equation}\label{eq:expression_j_layer_transfer_supplement}
\begin{aligned}
        \mathbf{T}^0=&\text{Diag}(\overbrace{\mathbf{H},\cdots,\mathbf{H}}^{(N+1) })\\
        &\times\text{Diag}(\left[
    \begin{array}{cc}
       0  & 1 \\
       1 & 0
    \end{array}
    \right],\underbrace{\mathbf{V},\cdots,\mathbf{V}}_{N },\left[
    \begin{array}{cc}
       0  & 1 \\
       1 & 0
    \end{array}
    \right])
\end{aligned}
\end{equation}
and that the considered $\theta$ of the vertical TBU at Row 1 and Column 0 occurs in the first $\mathbf{V}$ matrix inside $\text{Diag}(\cdot)$. Thus, we have:
\begin{equation}\label{eq:expression_deri_t0}
\begin{aligned}
        \frac{\partial \mathbf{T}^0}{\partial\theta}=&\text{Diag}(\overbrace{\mathbf{H},\cdots,\mathbf{H}}^{(N+1) })\\
        &\times\text{Diag}(\left[
    \begin{array}{cc}
       0  & 0 \\
       0 & 0
    \end{array}
    \right],\underbrace{\frac{\partial \mathbf{V}}{\partial \theta},\mathbf{0},\cdots,\mathbf{0}}_{N },\left[
    \begin{array}{cc}
       0  & 0 \\
       0 & 0
    \end{array}
    \right]) \; .
\end{aligned}
\end{equation}
Since ${\partial \mathbf{V}}/{\partial \theta}$ is straightforward using Eqs.~(~\ref{eq:relation_f_s}) and (\ref{eq:our_compact_S}) in the main text, we can calculate ${\partial\mathbf{a}_{M}^{(I)}}/{\partial \theta}$ by using Eqs. (\ref{eq:expression_deri_t0}), (\ref{eq:deri_tstar}), (\ref{eq:block_deri_tstar}), and (\ref{eq:deri_forward_output}) in sequence. With the derivative ${\partial\mathbf{a}_{M}^{(I)}}/{\partial \theta}$ available, it is easy to calculate the derivative of $Cost$ defined in Eq.~(\ref{eq:cost}) in the main text with respect to $\theta$:
\begin{equation}\label{eq:deri_cost}
    \frac{\partial Cost}{\partial \theta}=2\sum_{k=1}^{N_{\text{grid}}}\sum_{n=1}^N \left[a_{2n,M}^{(I)}(\omega_k)-U_n(\omega_k)\right]\frac{\partial a_{2n,M}^{(I)}(\omega_k)}{\partial\theta} \; .
\end{equation}
Here $\frac{\partial a_{2n,M}^{(I)}(\omega_k)}{\partial\theta}$ is the entry at index $2n$ of $\frac{\partial\mathbf{a}_{M}^{(I)}}{\partial \theta}$.

\subsection{\secrev{Color Cell Order in Heatmap}}\label{sec:appendix_added}

\begin{figure}[!htb]
    \centering
    \includegraphics[width=1.0\linewidth]{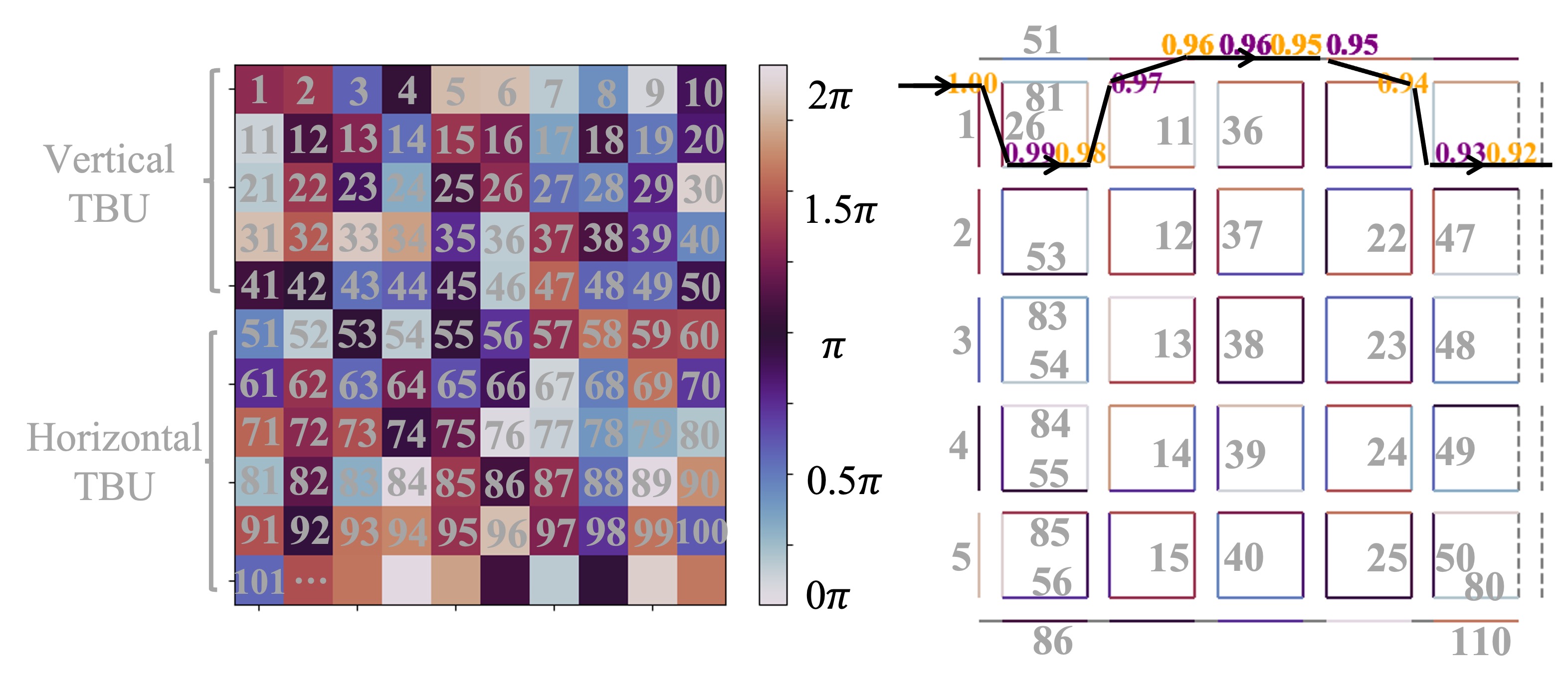}
\caption{\secrev{A demonstration of how we plot the heatmap using Fig.~\ref{fig:res_case1}~(c) and~(e) as an example.}}
    \label{fig:heatmap_demonstration}
\end{figure}

\secrev{We replot Fig.~\ref{fig:res_case1}~(c) and~(e) in  Fig.~\ref{fig:heatmap_demonstration} below to demonstrate how we draw the heatmap. Starting from the top left corner of the heatmap with a row-first order, the color cells represent $\theta$ of vertical TBUs (cells with gray number annotations 1 through 25), then $\phi$ of vertical TBUs (cells with annotations 26 through 50), then $\theta$ of horizontal TBUs (cells annotated from 51 to 80), and finally $\phi$ of horizontal TBUs (cells annotated from 81 to 110).}

\subsection{Power Coupling and Common Phase Shift}\label{sec:appendixc}

Recall the compact model of a TBU as shown in Eq.~(\ref{eq:form_of_Smatrix_wgs}). If we only focus on the first several terms, we have:
\begin{equation}\label{eq:power_common}
\begin{aligned}
     &   0.5\times\left[
    \begin{array}{cc}
        1 &  -\cj \\
        -\cj &  1
    \end{array}
    \right]
    \left[
    \begin{array}{cc}
        \e^{-\cj\theta} &  0 \\
        0 &  \e^{-\cj\phi}
    \end{array}
    \right]
    \left[
    \begin{array}{cc}
        1 &  -\cj \\
        -\cj &  1
    \end{array}
    \right] \\
    &= 0.5\left[
    \begin{array}{cc}
        \e^{-\cj\theta}-\e^{-\cj\phi} &  -\cj\e^{-\cj\theta}-\cj\e^{-\cj\phi} \\
        -\cj\e^{-\cj\theta}-\cj\e^{-\cj\phi} &   -\e^{-\cj\theta}+\e^{-\cj\phi}
    \end{array}\right] \\
    & = 0.5\e^{-\cj\phi}\left[
    \begin{array}{cc}
        \e^{-\cj(\theta-\phi)}-1 &  -\cj(\e^{-\cj(\theta-\phi)}+1) \\
        -\cj(\e^{-\cj(\theta-\phi)}+1) &   -\e^{-\cj(\theta-\phi)}+1
    \end{array}\right] \\
        & = \e^{-\cj\phi}\left[
    \begin{array}{cc}
        \cj\sin\frac{\alpha}{2}\,\e^{\cj\frac{\alpha}{2}} &  -\cj\cos\frac{\alpha}{2}\,\e^{\cj\frac{\alpha}{2}} \\
        -\cj\cos\frac{\alpha}{2}\,\e^{\cj\frac{\alpha}{2}} &   -\cj\sin\frac{\alpha}{2}\,\e^{\cj\frac{\alpha}{2}}
    \end{array}\right] \\
            & = \cj\e^{-\cj\phi}\e^{\cj\frac{\alpha}{2}}\left[
    \begin{array}{cc}
        \sin\frac{\alpha}{2}\, &  -\cos\frac{\alpha}{2} \\
        -\cos\frac{\alpha}{2} &   -\sin\frac{\alpha}{2}
    \end{array}\right] \\
                & = \underbrace{\e^{\cj(\frac{\pi}{2}-\phi+\frac{\alpha}{2})}}_{\text{Common phase shift}}\, \times \,\underbrace{\left[
    \begin{array}{cc}
        \sin\frac{\alpha}{2}\, &  -\cos\frac{\alpha}{2} \\
        -\cos\frac{\alpha}{2} &   -\sin\frac{\alpha}{2}
    \end{array}\right]}_{\text{Power coupling matrix}} \\
\end{aligned}
\end{equation}
where for simplicity, we have denoted $\alpha=-(\theta-\phi)$, and used the following equations to do the simplification:
\begin{equation}
    \begin{aligned}
         0.5(\e^{\cj\alpha}-1)&=\cj\sin\frac{\alpha}{2}\,\e^{\cj\frac{\alpha}{2}}\\
        0.5(\e^{\cj\alpha}+1)&=\cos\frac{\alpha}{2}\,\e^{\cj\frac{\alpha}{2}} \; .\\
    \end{aligned}
\end{equation}
The power coupling ratio is represented by $(\cos\frac{\alpha}{2})^2=\cos^2\frac{\phi-\theta}{2}$. This is shown by the shaded grey area in Fig.~\ref{fig:res_case1}~(f) of the main text. Based on Eq.~(\ref{eq:power_common}), we define $\frac{\pi}{2}-\phi+\frac{\alpha}{2}=\frac{\pi}{2}-\frac{\phi+\theta}{2}$ as the common phase shift and mark it using colors in the background of Fig.~\ref{fig:res_case1}~(f).

\subsection{\rev{Considering Thermal Cross-Talk and Other Non-Idealities}}\label{sec:appendixd}

\rev{Recall in the main text, we solve Eq.~(\ref{eq:cost}) to obtain a configuration for a desired light processing function. In one important real-world scenario (thermally controlled phase shifters), the PPIC delivers power to heaters at the phase shifters. However, this process involves non-ideal thermal cross-talk. This can be mathematically represented by~\cite{Maziyar2019canceling}:} 
\begin{equation}\label{eq:thermal_decomp}
    \rev{\mathbf{h}(\boldsymbol{\Phi} \mathbf{p})=\mathbf{x}}
\end{equation}
\rev{where $\boldsymbol{\Phi}$ is the thermal coefficient matrix, usually with ones on the diagonals and small off-diagonal entries (i.e., diagonal-dominant), $\mathbf{p}$ denotes the delivered power by the heaters at phase shifters, $\mathbf{h}$ represents the function mapping from power to phase shift, and $\mathbf{x}$ represents all phase shifts. As an example, an ideal case without thermal cross-talk corresponds to $\boldsymbol\Phi=\mathbf{I}$.}

\rev{With the help of Eq.~(\ref{eq:thermal_decomp}), our proposed method remains applicable when considering thermal cross-talk. Specifically, we can now optimize with respect to $\mathbf{p}$ instead of $\mathbf{x}$:}

\begin{equation}\label{eq:cost_thermal}
\begin{aligned}
    \rev{\min_{\mathbf{p}}} \quad &\rev{Cost=\sum_{k=1}^{N_{\text{grid}}}\sum_{n=1}^N \left|a_{2n,M}^{(I)}(\omega_k,\mathbf{x})-U_n(\omega_k)\right|^2} \\
    &\rev{s.t., \quad \mathbf{h}(\boldsymbol{\Phi} \mathbf{p})=\mathbf{x}}
\end{aligned}
\end{equation}
\rev{or more densely, }
\begin{equation}\label{eq:cost_thermal2}
    \rev{\min_{\mathbf{p}} \quad Cost=\sum_{k=1}^{N_{\text{grid}}}\sum_{n=1}^N \left|a_{2n,M}^{(I)}(\omega_k,\mathbf{h}(\boldsymbol{\Phi} \mathbf{p}))-U_n(\omega_k)\right|^2 \; .}
\end{equation}
\rev{Note that $\mathbf{h}$ and $\boldsymbol{\Phi}$ are known (or could be characterized) once a PPIC is fabricated. Thus, the above optimization is well-defined with respect to $\mathbf{p}$. Since the extra involved operations, multiplying by $\boldsymbol{\Phi}$ and mapping by $\mathbf{h}$, are differentiable, we can still obtain analytical gradients and adopt gradient descent optimization to solve Eq.~(\ref{eq:cost_thermal2}). A similar treatment can be adopted for other non-idealities.}

\subsection{\rev{The Compact Model Under Non-Ideality}}\label{sec:appendixe}

\rev{Our compact model shown in Eqs.~(\ref{eq:form_of_Smatrix}) and~(\ref{eq:our_compact_S}) assumes that the directional couplers achieve 50\%:50\% splitting; however, this is not the case in many real-world scenarios. The proposed compact model should be refined if non-idealities (e.g., process variation, non-ideal beam splitting) are important and need to be considered. Here we show how to refine the compact model for an example non-ideality:}

\begin{equation}\label{eq:form_of_Smatrix_process}
    \rev{\mathbf{F} = \frac{1}{2}\cdot\mathbf{M}(\eta_1)\cdot
    \left[
    \begin{array}{cc}
        \e^{-\cj\theta} &  0 \\
        0 &  \e^{-\cj\phi}
    \end{array}
    \right]\cdot \mathbf{M}(\eta_2)
    \cdot
    \alpha\e^{-\cj\omega\frac{n_{\text{eff}}L}{c}}}
\end{equation}
\rev{where $\{\eta_1,\eta_2\}$ represent the deviation of directional coupler from 50\%:50\%, and $\mathbf{M}(\eta)$ is defined as~\cite{Bandyopadhyay21forward}:}

\begin{equation}
    \rev{\mathbf{M}(\eta)=
    \left[
    \begin{array}{cc}
       \cos(\pi/4+\eta)  & -j\sin(\pi/4+\eta)  \\
        -j\sin(\pi/4+\eta) & \cos(\pi/4+\eta) \\
    \end{array}
    \right]}
\end{equation}
\rev{Note that $\{\eta_1,\eta_2,\alpha,L\}$ should be treated as random variables due to process variation. A commonly used assumption is that these random variables can be decomposed as the summation of determinisitic variables and perturbation variables:}
\begin{equation}
\rev{\begin{aligned}
        \eta_1=\eta_1^\star+\Delta\eta_1\\
        \eta_2=\eta_2^\star+\Delta\eta_2\\
        \alpha=\alpha^\star+\Delta\alpha\\
        L = L^\star + \Delta L\\
    \end{aligned}}
\end{equation}
\rev{where $\{\eta_1^\star,\eta_2^\star,\alpha^\star,L^\star\}$ denote the deterministic nominal design values, and $\{\Delta\eta_1,\Delta\eta_2,\Delta\alpha,\Delta L\}$ represent the probabilistic perturbation introduced in manufacturing. For different TBUs, they share the values of $\{\eta_1^\star,\eta_2^\star,\alpha^\star,L^\star\}$, but their $\{\Delta\eta_1,\Delta\eta_2,\Delta\alpha,\Delta L\}$ values are different. Moreover, we can further combine Eq.~(\ref{eq:form_of_Smatrix_process}) with Eq.~(\ref{eq:form_of_Smatrix_general_wgs}) to generate a more comprehensive compact model spanning all waveguides in the TBU.}

\rev{For any given instantiation of process variations (e.g., using random sampling), the compact model remains differentiable, and thus our proposed method can be used to generate a solution for that instance. Evaluation of performance degradation or yield (e.g., using Monte Carlo method), becomes possible. Future research might consider the robust synthesis problem, building on the method proposed here.}

\end{document}